\documentclass[twocolumn,pra,superscriptaddress]{revtex4}
\usepackage[latin9]{inputenc}
\usepackage{color}
\usepackage{bm}
\usepackage{amsmath}
\usepackage{amssymb}
\usepackage{bm}
\usepackage{braket}
\usepackage{graphicx}
\usepackage{tikz}
\usepackage{mathrsfs} % cal F fidelity
\usepackage{float}
\usepackage{bbm}
\usepackage{epstopdf}
\usepackage{epsfig}
\usepackage{verbatim}
\usepackage{array}
\usepackage{setspace}
\usepackage{times}
\usepackage{algorithm}
\usepackage{algpseudocode}
\usepackage{booktabs}
\usepackage{multirow}

\usepackage[unicode=true,
 bookmarks=true,bookmarksopen=false,
 breaklinks=false,pdfborder={0 0 1},backref=false,colorlinks=true]
 {hyperref}
\hypersetup{
 linkcolor=magenta, urlcolor=blue, citecolor=blue, pdfstartview={FitH}, hyperfootnotes=false, unicode=true}

\newcolumntype{C}[1]{>{\centering\arraybackslash$}p{#1}<{$}}

\begin{document}

\title{Automating Variational Quantum Sensing through Reinforcement-Learned Circuit Structures}

\author{Jie Liu}
\affiliation{Department of Physics, City University of Hong Kong, Tat Chee Avenue, Kowloon, Hong Kong SAR, China}
\affiliation{City University of Hong Kong Shenzhen Research Institute, Shenzhen, Guangdong 518057, China}
\affiliation{Quantum Science Center of Guangdong-Hong Kong-Macao Greater Bay Area, Shenzhen, Guangdong 518045, China}
\author{Xin Wang}
\email{x.wang@cityu.edu.hk}
\affiliation{Department of Physics, City University of Hong Kong, Tat Chee Avenue, Kowloon, Hong Kong SAR, China}
\affiliation{City University of Hong Kong Shenzhen Research Institute, Shenzhen, Guangdong 518057, China}

\date{\today}

%\begin{abstract}
%Variational quantum sensing optimizes probe states and measurements using parametrized quantum circuits, but existing approaches typically rely on fixed circuit architectures that constrain the accessible state manifold. We introduce \textsc{AutoQSense}, a reinforcement-learning-based framework that treats circuit structure itself as a trainable object, jointly optimizing preparation and measurement architectures to maximize Fisher information under hardware constraints. In a two-qubit benchmark with known optimal solutions, \textsc{AutoQSense} recovers the maximal quantum Fisher information achievable by a universal circuit, establishing expressive completeness of the learned architectures. Compared to fixed hardware-efficient ans\"atze, automated architecture search achieves higher precision and improved classical-to-quantum Fisher information attainability, particularly in noisy regimes. Structural analysis shows that the learned probes transition from GHZ-like states in the ideal limit to partially entangled, noise-adapted states as decoherence increases. To address scalability, we develop a distributed extension based on centralized training and decentralized execution, in which local agents optimize intra-block structures while a budgeted agent allocates inter-block entangling operations. This approach enables resource-aware discovery of sensing circuits in larger systems. Our results establish architecture learning as a systematic tool for adaptive and hardware-efficient quantum metrology.
%\end{abstract}

\begin{abstract}
Variational quantum sensing offers a promising route to high-precision parameter estimation, but its performance depends strongly on the circuit architectures used for probe preparation and measurement. Existing approaches typically optimize continuous parameters within predefined ans\"atze, restricting the accessible design space and limiting adaptation to sensing tasks and hardware constraints. Here, we introduce \textsc{AutoQSense}, a reinforcement-learning framework that searches circuit architectures using Fisher-information-based objectives. For few-qubit systems, a single agent sequentially constructs preparation and measurement circuits. For larger systems, a distributed formulation assigns local circuit design to subsystem agents and inter-block entanglement to a budgeted agent. Numerical results show that the learned architectures recover known benchmark strategies, adapt to dephasing noise, and outperform fixed hardware-efficient ans\"atze while using fewer entangling gates. These results establish \textsc{AutoQSense} as a resource-aware approach to adaptive and hardware-compatible quantum sensing.

\end{abstract}

\maketitle

\section{Introduction} \label{Intro}

Quantum sensing exploits unique quantum mechanical resources, such as coherence, entanglement, and squeezing, to enhance the precision of parameter estimation beyond classical limits \cite{Kitagawa1993squeeze,toth2014quantum,Degen2017sensing,Philipp2018ensemble,Pablo2024entanglement,Kannath2025squeeze}. By encoding an unknown parameter into a quantum probe and performing tailored measurements, quantum sensors can achieve sensitivities approaching the Heisenberg limit, surpassing the standard quantum limit attainable with classical resources alone \cite{demkowicz2012elusive}. This paradigm underpins a wide range of emerging quantum technologies, including atomic clocks \cite{ludlow2015atomic}, magnetometry \cite{budker2007optical}, gravitational-wave detection \cite{jia2024squeezing}, and nanoscale spectroscopy \cite{allert2022advances}. As quantum technologies continue to develop, the ability to systematically engineer high-performance probe states and measurement strategies has become a central challenge in quantum metrology and a key enabler of scalable quantum-enhanced devices.

The achievable precision in quantum parameter estimation is fundamentally limited by the quantum Cram\'er--Rao bound, which is governed by the quantum Fisher information (QFI) of the probe state with respect to the encoded parameter \cite{helstrom1969quantum}. The QFI therefore characterizes the maximum information that can, in principle, be extracted from the quantum state under an optimal measurement \cite{braunstein1994statistical,paris2009quantum,yu2022quantum}. In realistic experiments, the classical Fisher information (CFI) associated with the chosen measurement strategy quantifies the amount of parameter information that is actually accessible from the measurement outcomes \cite{braunstein1994statistical,paris2009quantum}. Consequently, the central design problem in quantum sensing is the joint optimization of state preparation and measurement in order to maximize QFI or CFI under realistic physical constraints. While optimal states can sometimes be derived analytically for simple models, closed-form solutions are generally unavailable in noisy, interacting, or many-body settings \cite{chabuda2020tensor}. This motivates the development of variational and numerical approaches for sensing protocol design.

Variational quantum sensing (VQS) has emerged as a powerful framework to address this challenge \cite{koczor2020variational,meyer2021variational,maclellan2024end,zuniga2024variational}. In VQS, parametrized quantum circuits (PQCs) are employed to generate probe states and, in some formulations, to implement measurement strategies. The circuit parameters are optimized to maximize a metrological objective function, typically the QFI or CFI. This approach offers two major advantages. First, it provides a hardware-compatible ansatz tailored to the native gate set of a given platform. Second, it transforms sensing design into an optimization problem amenable to gradient-based or gradient-free methods. Recent studies have demonstrated that variationally optimized circuits can outperform conventional sensing strategies in noisy and constrained settings, highlighting the promise of automated protocol discovery \cite{marciniak2022optimal,kaubruegger2023optimal,liao2024quantum}.

\begin{figure*}
	\includegraphics[width=\linewidth]{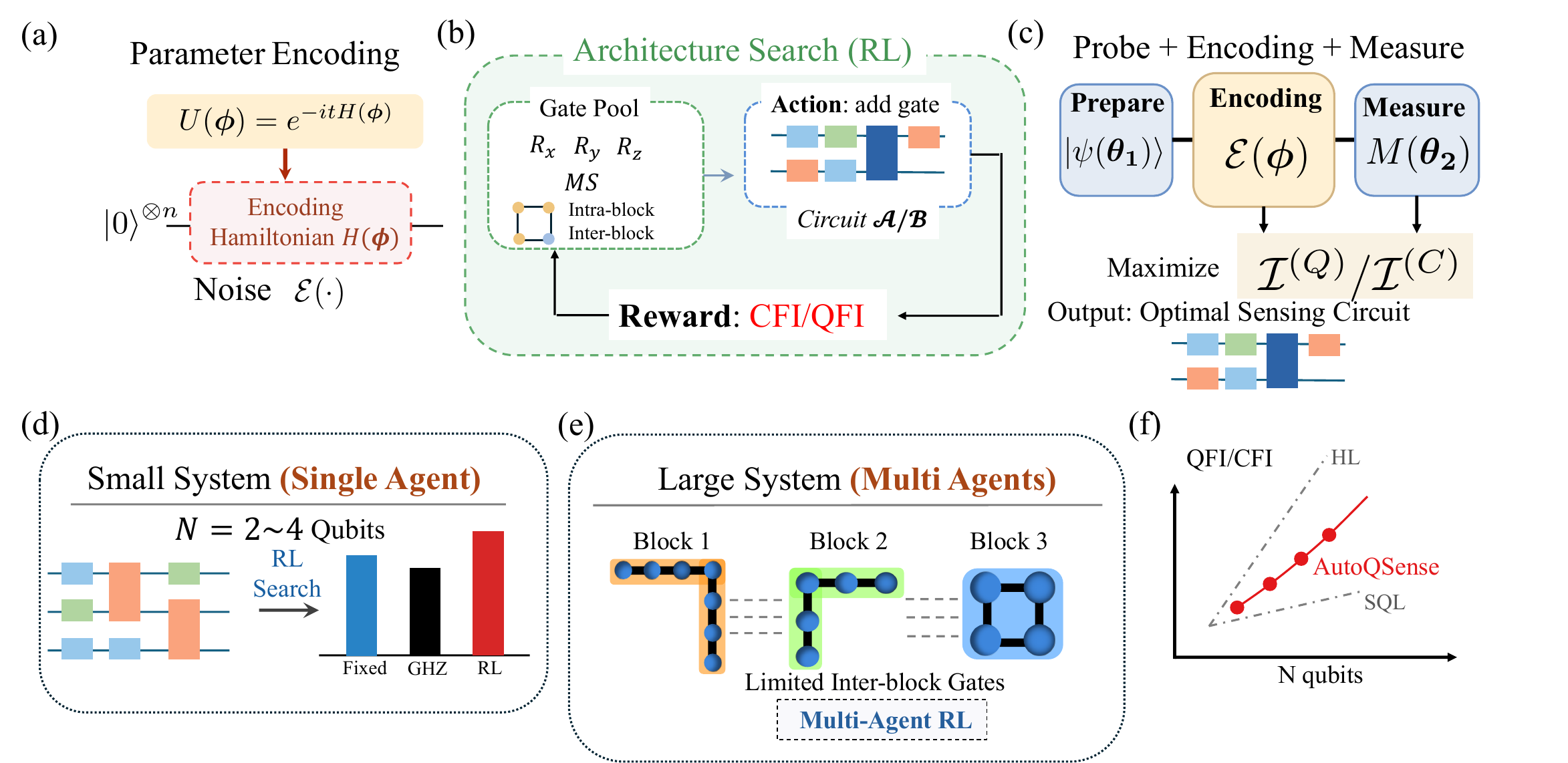}
	\caption{\textbf{Automating variational quantum sensing with reinforcement-learned circuit structures}. (a) An unknown parameter is encoded in an $n$-qubit probe through an encoding Hamiltonian, in the presence of noise. (b) The \textsc{AutoQSense} reinforcement-learning loop: an agent selects gates from a predefined pool to construct a circuit, which is evaluated by a CFI or QFI reward. (c) The resulting sensing protocol prepares a parameterized probe, encodes the unknown parameter, and applies a parameterized measurement to maximize the QFI or CFI. (d) For small systems, a single agent searches for an optimized circuit and compares its performance with fixed and GHZ-state baselines. (e) For larger systems, distributed-\textsc{AutoQSense} partitions the system into blocks controlled by multiple agents, with restricted inter-block gates. (f) The intended scaling is illustrated by the QFI or CFI as a function of qubit number, showing \textsc{AutoQSense} advancing beyond the standard quantum limit.}
	\label{fig:flowchart}
\end{figure*}

Despite these advances, nearly all existing VQS approaches rely on a \emph{fixed} circuit ansatz whose structure is chosen \emph{a priori}. Typically, one specifies a layered architecture composed of repeated blocks of single-qubit rotations and entangling gates, and only the continuous parameters are optimized. While such ans\"atze are convenient and hardware-friendly, they implicitly restrict the accessible state manifold and may exclude metrologically superior probe states outside this manifold. The choice of architecture therefore becomes a hidden hyperparameter that strongly influences performance. In many-body systems and noisy environments, the optimal structure is far less obvious, making fixed ans\"atze a significant constraint on achievable precision.
This observation highlights a key limitation of conventional variational approaches: parameter optimization cannot overcome architectural constraints on expressivity. In VQS, circuit structure must therefore be optimized as an integral part of the estimation strategy.

%This observation points to a more fundamental limitation in current variational approaches: optimizing circuit parameters alone is insufficient when the expressive capacity of the ansatz is itself constrained by its architecture. In quantum sensing, where metrological performance depends sensitively on global properties such as entanglement structure and parameter encoding, the choice of circuit topology becomes an intrinsic part of the estimation strategy rather than a secondary design choice.

Recent advances in quantum computing have begun to address this challenge by treating circuit construction as an optimization problem in its own right. In particular, quantum architecture search (QAS) has introduced algorithmic frameworks in which circuit structures are generated and refined automatically, rather than specified \emph{a priori} \cite{ye2021quantum,du2022quantum,zhang2022differentiable,huang2024adaptive,anastasiou2024tetris,grimsley2019adaptive}. 
These automated approaches, often based on reinforcement learning (RL) \cite{rapp2025reinforcement,zhang2021distributed,dai2024quantum}, evolutionary algorithms \cite{du2022quantum}, or gradient-based adaptive methods \cite{anastasiou2024tetris}, can identify architectures that outperform manually designed ans\"atze across a range of tasks.

%However, existing QAS frameworks are largely tailored to quantum optimization or quantum chemistry objectives, where performance is measured by energy. In contrast, sensing introduces a fundamentally different objective: under fixed parameter-encoding dynamics, the circuit must be designed to prepare probe states and measurement strategies that maximize the statistical distinguishability of the resulting measurement outcomes, rather than minimizing the statistical average value in quantum optimization or quantum chemistry. This shifts the design problem from approximating a target state to shaping the geometry of the quantum statistical model itself.

However, existing QAS frameworks have been developed primarily for quantum optimization and quantum chemistry, where circuit performance is typically evaluated through an energy or cost expectation value. Quantum sensing poses a distinct objective: under fixed parameter-encoding dynamics, the circuit must prepare probe states and implement measurements that maximize the distinguishability of parameter-dependent outcomes. This reframes architecture search as the task of engineering the geometry of the quantum statistical model to enhance parameter sensitivity.

In this work, we formulate sensing protocol design as a structure-learning problem driven by metrological performance. We introduce \textsc{AutoQSense}, an RL framework that uses the QFI or CFI to discover circuit architectures for probe-state preparation and measurement. The framework thus optimizes the input and readout structures surrounding fixed parameter-encoding dynamics, directly linking circuit structure to estimation performance. Fig.~\ref{fig:flowchart} provides an overview of the single-agent and distributed architecture-search workflows considered in this work.

For small systems, we cast circuit construction as a sequential decision process in which an agent selects gates from a predefined library. This allows the circuit topology to emerge adaptively, enabling the exploration of probe states and measurements beyond fixed parametrized ans\"atze.

For larger systems, we develop a multi-agent extension inspired by modular and distributed quantum architectures \cite{brown2016co,niu2023low,singh2025modular}. The system is partitioned into subsystems, each controlled by an agent that constructs a local circuit, while a separate agent introduces inter-block entangling gates through a restricted interface. This modular decomposition reduces the global search space while retaining the ability to generate multipartite correlations, enabling scalable circuit discovery through coordinated local policies.

Together, the single- and multi-agent formulations enable architecture search from few-qubit systems to larger modular settings under explicit resource and noise constraints. The learned circuits adapt to noise and can outperform fixed hardware-efficient ans\"atze while using fewer entangling gates. These results establish automated architecture discovery as a practical route toward adaptive, resource-efficient, and hardware-aware quantum sensing.

The remainder of this paper is organized as follows. 
In Sec.~\ref{sec:methods}, we introduce the \textsc{AutoQSense} framework, define the QFI and CFI objectives, and formulate \textsc{AutoQSense} as both single-agent and distributed multi-agent RL problems for circuit architecture discovery. 
In Sec.~\ref{sec:results}, we benchmark the proposed method in few-qubit and large-scale sensing tasks, comparing the learned architectures with standard protocols and fixed ans\"atze under resource and noise constraints. 
Finally, in Sec.~\ref{sec:conclusion}, we summarize the main findings and discuss the outlook for automated, adaptive, and hardware-aware quantum sensing.

\section{Method} \label{sec:methods}

Our approach combines VQS with RL-based architecture search in a nested optimization framework. For fixed preparation and measurement architectures, $\mathcal{A}$ and $\mathcal{B}$, an inner VQS loop optimizes the circuit parameters $\bm{\theta}_1$ and $\bm{\theta}_2$ using quantum differentiation and gradient-based updates. An outer RL loop samples and updates $\mathcal{A}$ and, when applicable, $\mathcal{B}$. Each sampled architecture undergoes a short inner-loop optimization, and its resulting metrological performance provides the reward for the policy update. Continuous parameters are therefore optimized for each candidate structure, while the architecture distribution is refined through policy gradients.

\subsection{Variational Quantum Sensing with Parametrized Preparation and Measurement}

We consider a general quantum sensing task in which a d-dimensional unknown parameter vector
$\bm{\phi}\in\mathbb{R}^d$ is encoded into an $n$-qubit probe through a parameter-dependent quantum channel
$\mathcal{E}_{\bm{\phi}}$ (unitary encoding as a special case). In VQS, both the probe
preparation and the measurement strategy can be automated using PQCs. We write the
probe preparation circuit as
\begin{equation}
    \ket{\psi(\bm{\theta}_1,\mathcal{A})}
    =
    U_{\mathrm{prep}}(\bm{\theta}_1,\mathcal{A})\ket{0}^{\otimes n},
\end{equation}
where $\bm{\theta}_1\in\mathbb{R}^{m_1}$ collects variational parameters and $\mathcal{A}$ specifies the preparation
architecture such as gate types, ordering, and connectivity. After encoding, the probe becomes
\begin{equation}
    \rho_{\bm{\phi}}(\bm{\theta}_1,\mathcal{A})
    =
    \mathcal{E}_{\bm{\phi}}
    \bigl(
        \ket{\psi(\bm{\theta}_1,\mathcal{A})}\bra{\psi(\bm{\theta}_1,\mathcal{A})}
    \bigr).
\end{equation}
To model a learnable measurement, we introduce a parametrized measurement circuit
\begin{equation}
    U_{\mathrm{meas}}(\bm{\theta}_2,\mathcal{B}),
\end{equation}
with parameters $\bm{\theta}_2\in\mathbb{R}^{m_2}$ and architecture $\mathcal{B}$. A common experimental realization is
to apply $U_{\mathrm{meas}}$ and then measure in the computational basis $\{\ket{x}\}$, i.e.\ using projectors
$\Pi_x=\ket{x}\bra{x}$. More generally, the elements of effective projectors induced by the measurement circuit are
\begin{equation}
    M_x(\bm{\theta}_2,\mathcal{B})
    =
    U_{\mathrm{meas}}^\dagger(\bm{\theta}_2,\mathcal{B})\,
    \Pi_x\,
    U_{\mathrm{meas}}(\bm{\theta}_2,\mathcal{B}),
\end{equation}
so that the outcome distribution is
\begin{equation}\label{eq:prob}
    p(x|\bm{\phi};\bm{\theta}_1,\bm{\theta}_2,\mathcal{A},\mathcal{B})
    =
    \mathrm{Tr}\!\left[
        M_x(\bm{\theta}_2,\mathcal{B})\,
        \rho_{\bm{\phi}}(\bm{\theta}_1,\mathcal{A})
    \right].
\end{equation}
For conciseness, we will often suppress the dependence on $(\mathcal{A},\mathcal{B})$ and write
$p(x|\bm{\phi};\bm{\theta}_1,\bm{\theta}_2)$.

\paragraph{Cram\'er--Rao bounds.}
Let $\hat{\bm{\phi}}$ be an unbiased estimator constructed from $\nu$ independent repetitions of the measurement.
The classical Cram\'er--Rao bound (CRB) states that the covariance matrix of $\hat{\bm{\phi}}$ obeys \cite{helstrom1969quantum}
\begin{equation}\label{eq:crb-multi}
    \mathrm{Cov}(\hat{\bm{\phi}})
    \;\succeq\;
    \frac{1}{\nu}\,
    \bigl(\mathcal{I}^{(C)}_M(\bm{\phi})\bigr)^{-1},
\end{equation}
where $\mathcal{I}^{(C)}_M(\bm{\phi})\in\mathbb{R}^{d\times d}$ is the CFI matrix
associated with the chosen measurement. For $d=1$, this reduces to \cite{helstrom1969quantum}
\begin{equation}
    \mathrm{Var}(\hat{\phi}) \ge \frac{1}{\nu \mathcal{I}^{(C)}(\phi)}.
\end{equation}

\paragraph{CFI matrix.}
Given the outcome probabilities in Eq.~\eqref{eq:prob}, the CFI matrix is defined by
\begin{equation}\label{eq:cfi-def}
    \mathcal{I}^{(C)}_M(\bm{\phi})_{ij}
    =
    \sum_x
    \frac{1}{p(x|\bm{\phi})}
    \frac{\partial p(x|\bm{\phi})}{\partial \phi_i}
    \frac{\partial p(x|\bm{\phi})}{\partial \phi_j},
\end{equation}
with the convention that terms with $p(x|\bm{\phi})=0$ are omitted.
The derivatives can be written explicitly using
\begin{equation}\label{eq:dpdphi}
    \frac{\partial p(x|\bm{\phi})}{\partial \phi_i}
    =
    \mathrm{Tr}\!\left[
        M_x(\bm{\theta}_2,\mathcal{B})\,
        \frac{\partial \rho_{\bm{\phi}}(\bm{\theta}_1,\mathcal{A})}{\partial \phi_i}
    \right].
\end{equation}

\paragraph{QFI and its relation to CFI.}
The QFI matrix $\mathcal{I}^{(Q)}_M(\bm{\phi})$ quantifies the maximum achievable information over
\emph{all} possible measurements and yields the quantum Cram\'er--Rao bound (QCRB)
\begin{equation}\label{eq:qcrb}
    \mathrm{Cov}(\hat{\bm{\phi}})
    \;\succeq\;
    \frac{1}{\nu}\,
    \bigl(\mathcal{I}^{(Q)}_M(\bm{\phi})\bigr)^{-1}.
\end{equation}
For $d=1$, this reduces to
\begin{equation}
    \mathrm{Var}(\hat{\phi})
    \geq
    \frac{1}{\nu\,\mathcal{I}^{(Q)}(\phi)} .
\end{equation}
A standard definition uses the symmetric logarithmic derivatives (SLDs) $\{L_i\}_{i=1}^d$ satisfying
\begin{equation}\label{eq:sld}
    \frac{\partial \rho_{\bm{\phi}}}{\partial \phi_i}
    =
    \frac{1}{2}\left(L_i\rho_{\bm{\phi}}+\rho_{\bm{\phi}}L_i\right),
\end{equation}
and
\begin{equation}\label{eq:qfi-def}
    \mathcal{I}^{(Q)}_M(\bm{\phi})_{ij}
    =
    \mathrm{Re}\Bigl[\mathrm{Tr}\bigl(\rho_{\bm{\phi}} L_i L_j\bigr)\Bigr].
\end{equation}
The QFI matrix can be calculated as follows:
\begin{equation}
\begin{split}
	[\mathcal{I}^{(Q)}_M]_{ij} &= 2 \, \mathrm{vec}[\partial_{\phi_i} \rho(\bm{\phi})]^\dagger 
[ \rho(\bm{\phi})^* \otimes I \\
&+ I \otimes \rho(\bm{\phi}) ]^+ 
\mathrm{vec}[\partial_{\phi_j} \rho(\bm{\phi})],
\end{split}
\end{equation}
where $\mathrm{vec}[\cdot]$ is the vectorization of a matrix, and the superscript $+$ denotes the pseudoinverse \cite{vsafranek2018simple}.

Crucially, for any fixed measurement $\{M_x\}$ one always has the operator inequality
\begin{equation}\label{eq:cfi-leq-qfi}
    \mathcal{I}^{(C)}_M(\bm{\phi})
    \;\preceq\;
    \mathcal{I}^{(\mathrm{Q})}_M(\bm{\phi}),
\end{equation}
i.e., the CFI matrix cannot exceed the QFI matrix. Equality can be attained when the measurement is chosen optimally.
In the single-parameter case, there always exists a POVM saturating the QCRB under mild regularity conditions, and
a measurement in the eigenbasis of the SLD achieves $\mathcal{I}^{(C)}=\mathcal{I}^{(Q)}$.
In the multi-parameter case, simultaneous saturation generally requires compatibility conditions (e.g., commuting SLDs),
and measurement optimization becomes nontrivial. This motivates the introduction of a variational measurement circuit
$U_{\mathrm{meas}}(\bm{\theta}_2,\mathcal{B})$ that directly targets the operationally relevant CFI. In experiments, the CFI matrix is directly accessible from measurement data, whereas the QFI matrix generally requires substantially greater experimental access, often involve state characterization or specialized measurements.

\paragraph{VQS loss functions.}
Since the CFI matrix is the quantity that directly determines the attainable error for the \emph{chosen} measurement, we take it
as the primary training objective and optionally optimize the QFI matrix as an upper bound. For single-parameter sensing, a natural
loss is
\begin{equation}
    \mathcal{L}(\bm{\theta}_1,\bm{\theta}_2;\phi)
    =
    -\mathcal{I}^{(C)}(\phi;\bm{\theta}_1,\bm{\theta}_2),
    \label{eq:CFI_loss}
\end{equation}
or equivalently the reciprocal (which corresponds to the CRB up to $\nu$),
$\mathcal{L}=1/\mathcal{I}^{(C)}$, to emphasize error scaling. For multi-parameter sensing, common scalarizations of the
CFI matrix include
\begin{align}
    \mathcal{L}_{\mathrm{A}} &= \mathrm{Tr}\!\left[\bigl(\mathcal{I}^{(C)}_M\bigr)^{-1}\right]
    \quad \text{(A-optimality \cite{optimality})}, \\
    \mathcal{L}_{\mathrm{D}} &= -\log\det\!\left(\mathcal{I}^{(C)}_M + \lambda I\right)
    \quad \text{(D-optimality \cite{optimality})}, \label{eq:CFIM_loss}
\end{align}
where $\lambda>0$ is a small regularizer used when $\mathcal{I}^{(C)}_M$ is ill-conditioned. In this study, we use Eq.~\eqref{eq:CFI_loss} in single-parameter sensing and Eq.~\eqref{eq:CFIM_loss} in multi-parameter sensing for numerical convenience.

\paragraph{Gradient evaluation for VQS.}
The VQS inner loop optimizes $\bm{\theta}=(\bm{\theta}_1,\bm{\theta}_2)$ for fixed architectures
$(\mathcal{A},\mathcal{B})$. Differentiating Eq.~\eqref{eq:cfi-def} yields
\begin{equation}\label{eq:cfi-grad-general}
\begin{split}
    \frac{\partial \mathcal{I}^{(C)}_M(\bm{\phi})_{ij}}{\partial \theta_\ell}
    =
    \sum_x
    [
        -\frac{1}{p_x^2}
        \frac{\partial p_x}{\partial \theta_\ell}
        \frac{\partial p_x}{\partial \phi_i}
        \frac{\partial p_x}{\partial \phi_j}
        \\
        +
        \frac{1}{p_x}
        \frac{\partial^2 p_x}{\partial \theta_\ell\,\partial \phi_i}
        \frac{\partial p_x}{\partial \phi_j}
        +
        \frac{1}{p_x}
        \frac{\partial p_x}{\partial \phi_i}
        \frac{\partial^2 p_x}{\partial \theta_\ell\,\partial \phi_j}
    ],
\end{split}
\end{equation}
where $p_x \equiv p(x|\bm{\phi};\bm{\theta}_1,\bm{\theta}_2)$.
In practice, we compute gradients of the chosen scalar loss $\mathcal{L}$ with respect to $\bm{\theta}$ using quantum
differentiation of the underlying probabilities. Concretely, for any circuit parameter $\theta_\ell$ entering through a
gate of the form $\exp(-i\theta_\ell P/2)$ with $P^2=I$, the parameter-shift rule gives
\begin{equation}\label{eq:param-shift-prob}
    \frac{\partial p_x}{\partial \theta_\ell}
    =
    \frac{
        p_x(\theta_\ell+\tfrac{\pi}{2}) - p_x(\theta_\ell-\tfrac{\pi}{2})
    }{2}.
\end{equation}
Derivatives with respect to $\bm{\phi}$ required in Eq.~\eqref{eq:cfi-def} can be obtained either analytically
(for known encodings) or by a finite-difference or shift rule when the encoding itself is implemented as a unitary gate
$\exp(-i\phi_i G_i)$ on hardware. With these ingredients, we estimate $\nabla_{\bm{\theta}}\mathcal{L}$ by simulating the quantum circuit and update
$\bm{\theta}$ using the Adam optimizer.

\paragraph{Inner-loop training procedure (baseline VQS).}
For fixed circuit architectures $(\mathcal{A},\mathcal{B})$, the standard VQS training loop iterates:
\begin{enumerate}
    \item Prepare $\ket{\psi(\bm{\theta}_1,\mathcal{A})}$, apply encoding $\mathcal{E}_{\bm{\phi}}$, apply readout
    $U_{\mathrm{meas}}(\bm{\theta}_2,\mathcal{B})$, and estimate probabilities $p(x|\bm{\phi})$.
    \item Compute $\mathcal{I}^{(C)}_M(\bm{\phi};\bm{\theta}_1,\bm{\theta}_2)$ via Eq.~\eqref{eq:cfi-def} and evaluate the
    scalar loss $\mathcal{L}$ via Eq.~\eqref{eq:CFI_loss} or Eq.~\eqref{eq:CFIM_loss}.
    \item Compute $\nabla_{\bm{\theta}}\mathcal{L}$ using the parameter-shift rule and analytic/shift derivatives for
    $\bm{\phi}$ as needed, then update $(\bm{\theta}_1,\bm{\theta}_2)$ with Adam.
\end{enumerate}
This VQS routine serves as the \emph{inner loop} of our automated architecture learning method.

\subsection{Reinforcement Learning for Architecture Optimization}

We now elevate the circuit architectures $\mathcal{A}$ and $\mathcal{B}$ 
to optimization variables. In our framework, an RL agent can sample 
preparation architectures $\mathcal{A}$ and measurement architectures $\mathcal{B}$, 
while the environment evaluates their metrological performance after a short inner-loop 
optimization of the variational parameters.

\paragraph{Architecture parametrization.}
Assuming the preparation circuit contains $L$ layers and $p$ native gate types, we represent the preparation architecture as
\begin{equation}
    \mathcal{A} = (a_1,\dots,a_L),
    \qquad
    a_j \in \{1,\dots,p\},
\end{equation}
where $a_j$ selects one of the $p$ native gate types at layer $j$. Hardware constraints can be incorporated by restricting the gate set and connectivity available at each layer, while $L$ sets the circuit depth. The measurement architecture $\mathcal{B}$ is represented in the same way when it is also included in the search.

To enable parameter reuse across sampled architectures, we define a parameter table
\begin{equation}
    \mathcal{T}^{(\textrm{prep})} \in \mathbb{R}^{p\times L},
\end{equation}
where each entry $\mathcal{T}^{(\mathrm{prep})}_{i,j}$ stores the variational parameter 
associated with applying the $i$-th gate type at layer $j$ in $U_{\mathrm{prep}}$.
If the measurement architecture is also searched, we introduce 
$\mathcal{T}^{(\mathrm{meas})}$ analogously.

For a sampled architecture $\mathcal{A}$, the instantiated parameter vector is
\begin{equation}
    \bm{\theta}_1(\mathcal{A})
    =
    \left(
        \mathcal{T}^{(\mathrm{prep})}_{a_1,1},
        \dots,
        \mathcal{T}^{(\mathrm{prep})}_{a_L,L}
    \right).
\end{equation}
During training, one entry of $\mathcal{T}^{(\mathrm{prep})}$ may be updated multiple times 
whenever the corresponding gate-layer pair appears in sampled architectures. The same rules apply to the architecture $\mathcal{B}$ and its corresponding parameter table $\mathcal{T}^{(\mathrm{meas})}$.

\paragraph{Policy over architectures.}

We define a stochastic policy $\pi_\alpha$ parametrized by $\alpha$, which 
generates preparation architectures according to
\begin{equation}
    P(\mathcal{A}|\alpha)
    =
    \prod_{j=1}^{L}
    \pi_\alpha(a_j|j).
\end{equation}
If the measurement architecture $\mathcal{B}$ is also optimized, a joint policy
$P(\mathcal{A},\mathcal{B}|\alpha)$ can be defined analogously.

\paragraph{Inner-loop evaluation of an architecture.}

We illustrate the optimization process using the preparation architecture $\mathcal A$ as an example; the same process is applied to the measurement architecture $\mathcal B$. For each sampled architecture $\mathcal{A}^{(i)}$ (and optionally $\mathcal{B}^{(i)}$), 
we instantiate the sensing experiment
\begin{equation}
    U_{\mathrm{prep}}\bigg(\bm{\theta}_1(\mathcal{A}^{(i)}),\mathcal{A}^{(i)}\bigg),
\end{equation}
compute the QFI-based loss
\begin{equation}
    \mathcal{L}\bigl(\mathcal{A}^{(i)},\bm{\theta}\bigr),
\end{equation}
and evaluate
\begin{equation}
    \nabla_{\bm{\theta}} 
    \mathcal{L}\bigl(\mathcal{A}^{(i)},\bm{\theta}\bigr)
\end{equation}
via the parameter-shift rule.

We perform several inner-loop updates of $\bm{\theta}$ using Adam 
and write the updated values back into the parameter table $\mathcal{T}^{(\mathrm{prep})}$.

\paragraph{Policy gradient with sampled reward.}

Let $R(\mathcal{A}^{(i)})$ denote the reward computed from the final Fisher-information-based loss 
after inner-loop updates for architecture $\mathcal{A}^{(i)}$. 
The policy objective is
\begin{equation}
    J(\alpha)
    =
    \mathbb{E}_{\mathcal{A}\sim\pi_\alpha}
    \bigl[
        R(\mathcal{A})
    \bigr].
\end{equation}
Using the score-function identity,
\begin{equation}
    \nabla_{\alpha} J(\alpha)
    =
    \mathbb{E}_{\mathcal{A}\sim\pi_\alpha}
    \Bigl[
        R(\mathcal{A})
        \nabla_{\alpha}\log P(\mathcal{A}|\alpha)
    \Bigr],
\end{equation}
which we approximate using a sampled batch:
\begin{equation}
    \nabla_{\alpha} J(\alpha)
    \approx
    \frac{1}{N}
    \sum_{i=1}^{N}
    R\bigl(\mathcal{A}^{(i)}\bigr)
    \nabla_{\alpha}
    \log P\bigl(\mathcal{A}^{(i)}|\alpha\bigr),
\end{equation}
we then update the parameters $\alpha$ using the Adam optimizer with the gradient $\nabla_{\alpha} J(\alpha)$.

\paragraph{Fine-tuning after policy convergence.}

After convergence of the policy, we select the most probable architecture
\begin{equation}
    \mathcal{A}^{(\mathrm{max})}
    =
    \arg\max_{\mathcal{A}} P(\mathcal{A}|\alpha),
\end{equation}
and perform additional inner-loop optimization of the circuit parameters $\bm{\theta}_1$ until saturation. 
\paragraph{Hybrid optimization loop.}

Here we summarize the training loop of our method. Each training iteration proceeds as follows:
\begin{enumerate}
    \item Sample a batch of architectures 
    $\{\mathcal{A}^{(i)}\}_{i=1}^N \sim \pi_\alpha$.
    \item For each $\mathcal{A}^{(i)}$, perform several inner-loop updates 
    of $\bm{\theta}$ and update the parameter table.
    \item Compute rewards $R(\mathcal{A}^{(i)})$ from the final Fisher-information-based losses.
    \item Estimate $\nabla_\alpha J(\alpha)$ and update $\alpha$.
\end{enumerate}

The complete procedure is presented in Algorithm~\ref{alg:autoqsense}. This hybrid scheme combines quantum gradients for variational parameters 
with policy gradients for architectural exploration. By jointly adapting 
$\mathcal{A}$ (and optionally $\mathcal{B}$) together with 
$\bm{\theta}_1$ (and $\bm{\theta}_2$), \textsc{AutoQSense} 
automatically discovers sensing circuit structures that are tailored 
to the signal model and experimental constraints.

\begin{algorithm}[b]
\caption{Single-Agent \textsc{AutoQSense}}
\label{alg:autoqsense}
\begin{algorithmic}[1]
\Require Gate library $\mathcal{G}$, circuit depth $L$, policy parameters $\alpha$, parameter tables $\mathcal{T}^{(\mathrm{prep})}, \mathcal{T}^{(\mathrm{meas})}$
\Require Number of architectures per batch $N$, inner-loop steps $K$
\Ensure Optimized architecture $\mathcal{A}^\star$ and circuit parameters $(\bm{\theta}_1,\bm{\theta}_2)$

\For{each training iteration}
    \State Sample architectures $\{\mathcal{A}^{(i)}\}_{i=1}^N \sim \pi_\alpha$
    \For{each architecture $\mathcal{A}^{(i)}$}
        \State Instantiate parameters $\bm{\theta}^{(i)}$ from parameter tables
        \For{$k=1$ to $K$}
            \State Execute sensing circuit and estimate probabilities $p(x|\phi)$
            \State Compute Fisher-information-based loss $\mathcal{L}$
            \State Update $\bm{\theta}^{(i)}$ via parameter-shift gradients
        \EndFor
        \State Write updated parameters back to $\mathcal{T}$
        \State Compute reward $R(\mathcal{A}^{(i)})$
    \EndFor
    \State Update policy parameters $\alpha$ using policy gradients
\EndFor
\State Select most probable architecture $\mathcal{A}^\star$
\State Fine-tune $\bm{\theta}_1$ for fixed $\mathcal{A}^\star$
\end{algorithmic}
\end{algorithm}

\subsection{Distributed Reinforcement Learning via Centralized Evaluation and Decentralized Policies for Large-Scale Systems}

The architecture search in \textsc{AutoQSense} becomes increasingly challenging as the number of qubits grows. For an $n$-qubit circuit with depth $L$ and $p$ native gate types per layer, the number of candidate preparation structures $\mathcal{A}$ alone grows exponentially with circuit depth, and each candidate must be evaluated through repeated circuit executions to estimate the sensing objective. In quantum sensing, this evaluation is particularly expensive because the reward is not a direct property of the circuit: it must be inferred from measurement statistics and derivatives of the observed distribution with respect to the unknown parameter. This motivates a distributed formulation that preserves enough global expressivity while reducing the effective search complexity. Such a formulation is naturally aligned with the rapid development of modular and distributed quantum hardware, where large-scale devices are increasingly expected to be built from smaller, locally controlled quantum modules connected through limited inter-module couplings \cite{brown2016co,niu2023low,singh2025modular}.

This perspective is also consistent with recent advances in distributed quantum sensing, where spatially separated sensors are typically coordinated through shared entangled resources, quantum networks, or correlated measurements to estimate global functions of locally encoded parameters \cite{zhang2021distributed,guo2020distributed,bate2025experimental}. In such protocols, the sensing nodes often interact only locally with their respective signals, while nonlocal correlations are supplied through entangled probe states or joint measurement strategies. Therefore, a modular search strategy that separates local circuit construction from controlled inter-block entanglement is well matched to the structure of distributed sensing platforms.

\paragraph{Block partition with a dedicated inter-block agent.}
We partition the $n$-qubit system into $M$ disjoint blocks,
\begin{equation}
\mathcal{H}=\bigotimes_{m=1}^M \mathcal{H}_m,
\end{equation}
where block $m$ contains $n_m$ qubits and $\sum_{m=1}^M n_m=n$.
In distributed-\textsc{AutoQSense}, the preparation architecture $\mathcal{A}$ is constructed layer by layer from two interacting components:
(i) \emph{intra-block actions} selected by $M$ local agents, and
(ii) \emph{inter-block entangling actions} selected by an additional \emph{inter-block agent}.

Concretely, at each layer $\ell\in\{1,\dots,L\}$, local agent $m$ selects an intra-block gate pattern for its block, denoted by $a_{m,\ell}$, which specifies the portion of $\mathcal{A}$ acting on $\mathcal{H}_m$ at that layer. The number of gates applied by each local agent within a layer is also a tunable hyperparameter. In addition, after the local choices at layer $\ell$ are made, the inter-block agent determines whether and where to apply an entangling operation between blocks at that layer. We denote this inter-block decision by $e_{\ell}\in\mathcal{E}\cup\{\varnothing_{ij}\}$, where $\varnothing_{ij}$ is the null action between the $i$-th and $j$-th blocks (no inter-block gate) and $\mathcal{E}$ is a predefined set of allowed inter-block entangling primitives, e.g., fixed-connectivity two-qubit patterns across block boundaries.

To reflect realistic hardware constraints, where long-range entangling operations are costly, the inter-block agent is allowed to take at most $J$ non-null inter-block actions across the entire circuit. Formally, let $\mathbb{I}[e_\ell\neq\varnothing]$
be the indicator of whether an entangling action is used at layer $\ell$. Then the inter-block action budget constraint is
\begin{equation}\label{eq:entangle-budget}
    \sum_{\ell=1}^{L} \mathbb{I}[e_\ell\neq\varnothing] \le J.
\end{equation}
Once the budget is exhausted, the inter-block agent is forced to output $\varnothing$ for all remaining layers. This design enforces a physically meaningful inductive bias: the algorithm may employ entanglement as a scarce resource, deploying it only when it yields a tangible improvement in sensing performance.

With these definitions, the preparation circuit can be written as an ordered product over layers,
\begin{equation}\label{eq:prep-layerwise}
    U_{\mathrm{prep}}(\bm{\theta}_1,\mathcal{A})
    =
    \prod_{\ell=1}^{L}
    \Bigl[
        U_{\mathrm{int}}\bigl(\theta_\ell, e_\ell\bigr)
        \;
        \bigotimes_{m=1}^{M}
        U_{\mathrm{prep}}^{(m)}\bigl(\theta_{1,m,\ell}, a_{m,\ell}\bigr)
    \Bigr],
\end{equation}
where $U_{\mathrm{prep}}^{(m)}(\cdot)$ denotes the intra-block operation on block $m$ at layer $\ell$,
and $U_{\mathrm{int}}(\theta_\ell,e_\ell)$ denotes the inter-block entangling gate specified by $e_\ell$ (with parameter $\theta_\ell$ if applicable), with the convention that $U_{\mathrm{int}}(\theta_\ell,\varnothing)=I$.
The full architecture $\mathcal{A}$ is therefore the collection of intra-block selections $\{a_{m,\ell}\}$ together with the inter-block selections $\{e_\ell\}$.

\begin{figure*}[t]
	\includegraphics[width=\textwidth]{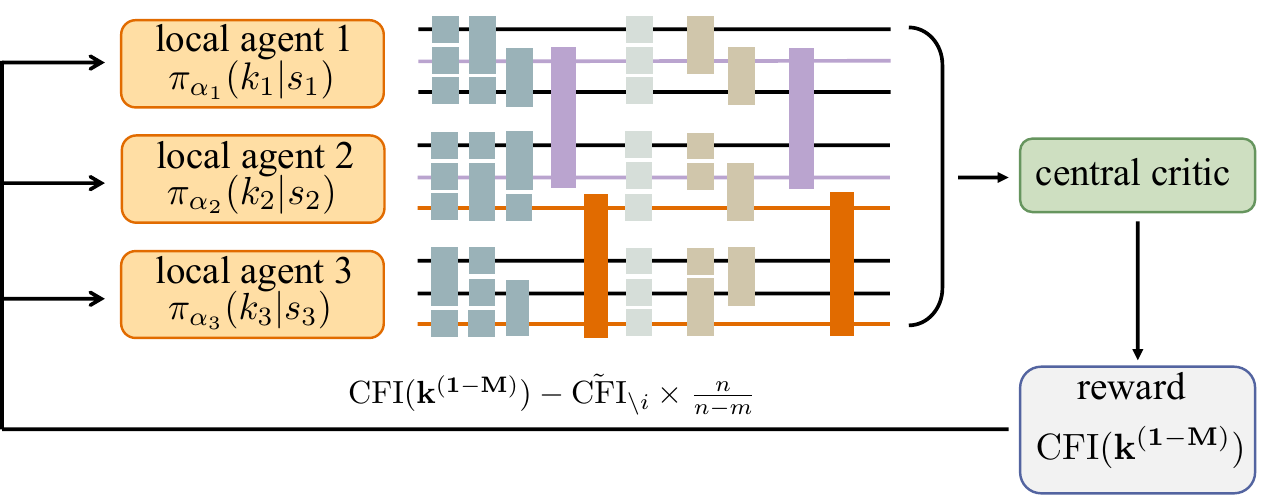}
	\caption{Schematic of the multi-agent RL framework for distributed quantum-sensor architecture search. Each local agent observes its local state $s_i$ and samples a local circuit action $k_i$ from its policy $\pi_{\alpha_i}(k_i|s_i)$. The local actions are assembled into a global circuit architecture $\bm{k}^{(1-M)}$, which is evaluated by a centralized evaluator. The resulting CFI, together with the estimated local contribution, is used as the reward to update the local policies.}
	\label{fig:dis_RL}
\end{figure*}

\paragraph{CTDE: centralized evaluation with decentralized policies.}
We adopt centralized training and decentralized execution (CTDE) to coordinate the $M+1$ agents \cite{amato2024introduction}. Each local agent $m$ maintains a policy $\pi_{\alpha_m}$ and the inter-block agent maintains a policy $\pi_{\alpha_{\mathrm{int}}}$. During execution, the joint policy factorizes as
\begin{equation}\label{eq:ctde-factor}
P(\mathcal{A}\mid\boldsymbol{\alpha})
=
\prod_{\ell=1}^{L}
\left[
\left(
\prod_{m=1}^{M}
\pi_{\alpha_m}(a_{m,\ell})
\right)
\pi_{\alpha_{\mathrm{int}}}(e_\ell)
\right],
\end{equation}
subject to the budget constraint in Eq.~\eqref{eq:entangle-budget}.
A centralized evaluator has access to the global trajectory (including all actions and the resulting sensing performance) and provides training signals to all agents. As illustrated in Fig.~\ref{fig:dis_RL}, the local actions are assembled into a global circuit, whose centralized evaluation supplies the reward signal used to update the local policies.

\paragraph{Reward design: CFI-only objective and physical motivation.}
In the distributed setting, we base rewards on the CFI because it is directly computable from measurement statistics and remains operationally meaningful at scale. In contrast, estimating the QFI (or QFI matrix) generally requires state characterization or access to optimal measurements, which becomes infeasible beyond small $n$. With the measurement circuit $U_{\mathrm{meas}}(\bm{\theta}_2,\mathcal{B})$
similarly sampled from the policy, executing the full sensing protocol yields outcome probabilities
\begin{equation}
p(x|\bm{\phi};\bm{\theta}_1,\bm{\theta}_2,\mathcal{A},\mathcal{B}),
\end{equation}
from which we compute the CFI matrix using Eq.~\eqref{eq:cfi-def}.
We define the global reward as
\begin{equation}\label{eq:global-reward}
    r_{\mathrm{global}}
    =
    -\mathcal{L}_D(\phi;\bm{\theta}_1,\bm{\theta}_2,\mathcal{A},\mathcal{B}).
\end{equation}
This choice is physically transparent: maximizing $r_{\mathrm{global}}$ improves the implemented measurement according to a D-optimal Fisher-information criterion.

\paragraph{Budgeted entanglement as a sparse resource.}
The inter-block agent's budget $J$ introduces an explicit notion of \emph{sparse entanglement}. Physically, inter-block entangling gates are typically the most error-prone and resource-intensive operations. Allowing the inter-block agent to place at most $J$ such operations forces the learned strategy to (i) rely primarily on robust intra-block optimization and (ii) use inter-block entanglement only when it produces a net metrological gain. This induces a structured trade-off between expressivity (ability to create global correlations) and practicality (entanglement cost), and empirically improves sample efficiency by constraining the outer-loop search.

\paragraph{Credit assignment via counterfactual difference rewards.}

A remaining challenge is credit assignment: the global reward $r_{\mathrm{global}}$ depends on the joint actions of all agents and is only revealed after completing the entire circuit, resulting in a delayed, global signal. As a result, it becomes difficult to disentangle the contribution of individual agents or local decisions to the overall performance, leading to poorer convergence during training. In generic multi-agent reinforcement learning, this issue is often addressed by introducing an additional learnable value network, such as QMIX \cite{rashid2020weighted} or related architectures, that factorizes the global reward into agent-wise components. While effective in many settings, such approaches rely on auxiliary neural networks that serve as black-box credit allocators and may obscure the relationship between individual actions and task performance.

In contrast, our credit assignment mechanism is constructed directly from the metrological objective and is designed to remain computationally tractable in large systems. Conceptually, for each local agent $m$ one may define a counterfactual architecture $\mathcal{A}_{\setminus m}$ by replacing that agent's intra-block operations with a null (identity) action, while keeping all other agents' decisions, including inter-block entanglement, unchanged. Evaluating the CFI matrix associated with $\mathcal{A}_{\setminus m}$ can provide a direct estimate of agent $m$'s marginal contribution to sensing performance. However, computing this quantity exactly would require executing the modified circuit and re-estimating measurement statistics for each agent-specific counterfactual, introducing a prohibitive overhead that scales linearly with the number of agents and circuit evaluations.

To avoid this additional cost, we employ a computationally inexpensive proxy that approximates the counterfactual CFI matrix without requiring further quantum executions. Specifically, we approximate the effect of removing agent $m$'s contribution by classically marginalizing over its measurement outcomes in the observed joint probability distribution. 
For a local block $m$, we define the marginalized distribution
\begin{equation}\label{eq:marginalize}
    \tilde{p}_{\setminus m}(x_{\setminus m}|\phi)
    =
    \sum_{x_m} p(x_m,x_{\setminus m}|\phi),
\end{equation}
from which we compute an approximate counterfactual CFI, denoted $\tilde{\mathcal{I}}^{(\mathrm{C})}_{\setminus m}$, using Eq.~\eqref{eq:cfi-def}. 
To quantify the contribution of the inter-block agent, we first compute, for each subsystem $m$, a local proxy CFI from the marginal measurement distribution
\begin{equation}
    p_m(\mathbf{x}_m|\boldsymbol{\theta})
    =
    \sum_{\mathbf{x}_{\bar m}}
    p(\mathbf{x}_m,\mathbf{x}_{\bar m}|\boldsymbol{\theta}),
\end{equation}
where $\mathbf{x}_{\bar m}$ denotes measurement outcomes outside subsystem $m$.
The corresponding local CFI score is
\begin{equation}
    F_m^{\mathrm{loc}}
    =
    \log\det\!\left(\mathcal{I}^{(C)}(p_m) + \lambda I\right)
\end{equation}.
To compare local and global sensing performance on the same scale, we define a
size-normalized local proxy
\begin{equation}
    F_{\mathrm{proxy}}^{\mathrm{loc}}
    =
    \frac{1}{M}
    \sum_{m=1}^{M}
    \frac{n}{n_m}
    F_m^{\mathrm{loc}},
\end{equation}
where $n_m$ is the number of qubits in subsystem $m$ and $n$ is the total number
of qubits. This quantity estimates the global CFI expected from the average
local information density, ignoring inter-block correlations.

We emphasize that this marginalization-based procedure does not reproduce the CFI matrix of the physically modified circuit in which gates are explicitly removed. 
Rather, it provides a low-cost proxy that captures how much information about the parameter $\phi$ is accessible from the remaining subsystems, while preserving a direct operational connection to the Fisher information. 

Using this proxy, we define the agent-specific reward for each local agent
\begin{equation}\label{eq:diff-reward}
    \tilde{r}_m
    =
    r_{\mathrm{global}}
    -
    \gamma_m\,\tilde{F}_{\setminus m},
\end{equation}
where $\gamma_m$ is a normalization factor introduced to remove the trivial size dependence expected under standard-quantum-limit scaling and $\tilde{F}_{\setminus m}=\log\det\!\left(\tilde{\mathcal{I}}_{\setminus m}^{(C)} + \lambda I\right)$. A natural choice is
\begin{equation}
	\gamma_m = \frac{n}{n-n_m},
\end{equation}
reflecting the linear scaling of Fisher information with the number of independent qubits under the standard quantum limit. 
With this normalization, $\tilde{r}_m>0$ indicates that block $m$ contributes a metrological gain beyond what would be expected from merely adding independent sensing resources. For the entangling agent, the reward $\tilde{r}_\mathrm{int}$ is defined as 
\begin{equation}
	\tilde{r}_\mathrm{int} = r_\mathrm{global}-F_{\mathrm{proxy}}^{\mathrm{loc}}
\end{equation}.

\paragraph{Policy updates.}
Each agent is trained with policy gradients using its proxy reward. For local agent $m$,
\begin{equation}\label{eq:pg-local}
    \nabla_{\alpha_m} J_m
    =
    \mathbb{E}
    \Biggl[
        \sum_{\ell=1}^{L}
        \tilde{r}_m\,
        \nabla_{\alpha_m}\log \pi_{\alpha_m}(a_{m,\ell}\,|\,s_{m,\ell})
    \Biggr],
\end{equation}
and similarly for the inter-block agent with reward $\tilde{r}_{\mathrm{int}}$. This CTDE formulation enables decentralized architectural decisions at execution time, while centralized evaluation and counterfactual rewards stabilize learning and improve sample efficiency. The distributed-\textsc{AutoQSense} algorithm flow is formally presented in Algorithm~\ref{alg:distributed_autoqsense}.

\begin{algorithm}[t]
\caption{Distributed \textsc{AutoQSense} (CTDE)}
\label{alg:distributed_autoqsense}
\begin{algorithmic}[1]
\Require Partition $\{\mathcal{H}_m\}_{m=1}^M$, entanglement budget $J$
\Require Local policies $\{\pi_{\alpha_m}\}$, inter-block policy $\pi_{\alpha_{\mathrm{int}}}$
\Require Parameter tables $\{\mathcal{T}^{(m)}\}$, inner-loop steps $K$
\Ensure Optimized distributed architecture $\mathcal{A}^\star$

\For{each training episode}
    \State Initialize remaining entanglement budget $J_{\mathrm{rem}} \gets J$
    \For{layer $\ell = 1$ to $L$}
        \For{each block $m$}
            \State Sample intra-block action $a_{m,\ell} \sim \pi_{\alpha_m}(\cdot)$
        \EndFor
        \If{$J_{\mathrm{rem}} > 0$}
            \State Sample inter-block action $e_\ell \sim \pi_{\alpha_{\mathrm{int}}}(\cdot)$
            \If{$e_\ell \neq \varnothing$}
                \State $J_{\mathrm{rem}} \gets J_{\mathrm{rem}} - 1$
            \EndIf
        \Else
            \State $e_\ell \gets \varnothing$
        \EndIf
    \EndFor
    \State Assemble full circuit architecture $\mathcal{A}$
    \State Perform inner-loop VQS optimization for $K$ steps
    \State Compute global reward $r_{\mathrm{global}}$ from CFI
    \State Compute proxy rewards $\{\tilde r_m\}$ via counterfactual marginalization
    \State Update all policies using policy gradients (CTDE)
\EndFor
\State Select most probable distributed architecture $\mathcal{A}^\star$
\end{algorithmic}
\end{algorithm}

Overall, distributed-\textsc{AutoQSense} preserves the physical structure of realistic platforms, namely dominant local control with scarce long-range entanglement, while enabling scalable architecture discovery. The explicit budgeted inter-block agent enforces that global correlations are introduced only when they are metrologically justified, yielding an interpretable mechanism for trading entangling resources against achievable sensing precision.

\section{Results} \label{sec:results}
%We now present numerical studies to evaluate the performance, robustness, and scalability of \textsc{AutoQSense}. The results are organized into two complementary regimes that reflect the core capabilities of the proposed framework. We first consider small-scale systems, where the full circuit expressivity can be systematically explored and, in certain cases, optimal strategies are analytically or numerically accessible. This setting enables controlled validation of the learned architectures, including comparisons against the universal two-qubit PQC, standard sensing protocols such as Ramsey and GHZ-based schemes under both noiseless and noisy conditions, showing how the discovered architecture recovers the GHZ strategy in the ideal limit and adapts toward partially entangled structures in the presence of noise, and the fixed-structure variational ans\"atze to assess the advantage of architecture search over fixed-structure variational design.
%We then turn to larger-scale systems, where exhaustive search is no longer feasible and circuit design must account for both computational and hardware constraints. In this regime, we employ the multi-agent variant of \textsc{AutoQSense} and benchmark its performance against hardware-efficient ans\"atze beyond 12 qubits. We further investigate how different subsystem partition strategies influence the resulting CFI, providing insight into the role of modular structure in scalable sensing protocol design.
We evaluate \textsc{AutoQSense} along three aspects: expressivity, noise adaptivity, and scalability. First, in few-qubit systems, we test whether RL-learned circuit structures can recover known optimal or near-optimal sensing strategies. We compare against a universal two-qubit benchmark, GHZ and Ramsey protocols, and fixed hardware-efficient ans\"atze. These benchmarks isolate the role of architecture search from ordinary continuous parameter optimization.

We then move to larger systems, where exhaustive architecture search becomes impractical. In this regime, we use the distributed multi-agent formulation of \textsc{AutoQSense} to decompose the global circuit-design problem into local intra-block decisions and budgeted inter-block entangling operations. This allows us to examine whether RL-learned architectures can improve Fisher information while reducing the number of two-qubit gates relative to fixed hardware-efficient circuits.

\subsection{Few-qubit benchmarks: validation and noise-adaptive sensing}

\subsubsection{Recovery of the universal two-qubit optimum}\label{sec:twoq_uni}
\begin{figure}
	\includegraphics[width=\linewidth]{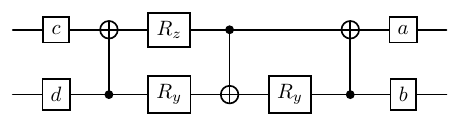}
	\caption{Universal two-qubit circuit. Here, $a$, $b$, $c$, and $d$ denote general $\mathrm{SU}(2)$ rotations.}
	\label{fig:two_qubit_circ}
\end{figure}

We first validate \textsc{AutoQSense} in the minimal nontrivial setting of a two-qubit sensing task, where a universal two-qubit PQC can be used to construct an upper-bound benchmark for the achievable sensitivity. This controlled setting serves two purposes. First, it allows us to verify that restricting the search to reinforcement-learned discrete architectures does not compromise the ability to reach the benchmark sensitivity. Second, it enables us to quantify how the gate budget controls the expressive capacity of \textsc{AutoQSense} and determines its ability to recover the optimal $\mathrm{Tr}(\mathcal{I}^{(Q)}_M)$.

\paragraph{Benchmark: universal two-qubit circuit.}

For two qubits, any unitary operation in $\mathrm{SU}(4)$ can be decomposed into a universal circuit consisting of single-qubit rotations and a finite number of entangling gates, as shown in Fig.~\ref{fig:two_qubit_circ}. Such a universal construction provides access to the entire two-qubit state manifold. We therefore use this sufficiently expressive universal two-qubit circuit as a reference benchmark, and compute the maximal achievable QFI matrix for the given encoding Hamiltonian by variationally optimizing all its parameters. We consider a multi-parameter estimation task involving two qubits subjected to a static magnetic field in three spatial directions. The dynamics of each qubit are governed by the Hamiltonian
\begin{equation}
	H^{(i)}(\bm{\phi}) =  \phi_x \sigma_x^{(i)} + \phi_y \sigma_y^{(i)} + \phi_z \sigma_z^{(i)},
\end{equation}
where $ \bm{\phi} = (\phi_x, \phi_y, \phi_z) $ are the unknown field amplitudes to be estimated. In this experiment, we set $ \bm{\phi} = (\pi/4, \pi/6, \pi/4) $. The encoding process is implemented by a unitary evolution followed by a dephasing channel:
\begin{equation}
	\rho(\bm{\phi})=\mathcal{E}(U_{\bm{\phi}}\rho U^\dag_{\bm{\phi}})= \sum_k K_k U_{\bm{\phi}}\rho U^\dag_{\bm{\phi}} K_k^\dag,
\end{equation}
with $U_{\bm{\phi}}=\exp(-iH(\bm{\phi})t)$ and fixed duration $ t = 1 $. The Kraus operators for the dephasing channel are defined as follows:
\begin{equation}
K_1 = \begin{pmatrix} \sqrt{1-\lambda} & 0 \\ 0 & 1 \end{pmatrix}, K_2 = \begin{pmatrix} \sqrt{\lambda} & 0 \\ 0 & 0 \end{pmatrix},
\label{eq:dephasing}
\end{equation}
where $\lambda$ denotes the noise strength and is set to $\lambda=0.05$ in the numerical simulations. This benchmark establishes an upper bound on the value of $\mathrm{Tr}(\mathcal{I}^{(Q)}_M)$ achievable by any unrestricted circuit.

\paragraph{Gate-budget-controlled architecture search.}

We now apply \textsc{AutoQSense} with varying gate budgets $L$. Specifically, we constrain the preparation architecture $\mathcal{A}$ to a fixed maximum number of layers $L$, and restrict the available gate types to a native gate set compatible with the universal construction. For each budget setting, \textsc{AutoQSense} optimizes the circuit structure and parameters.

\begin{figure}
	\includegraphics[width=\linewidth]{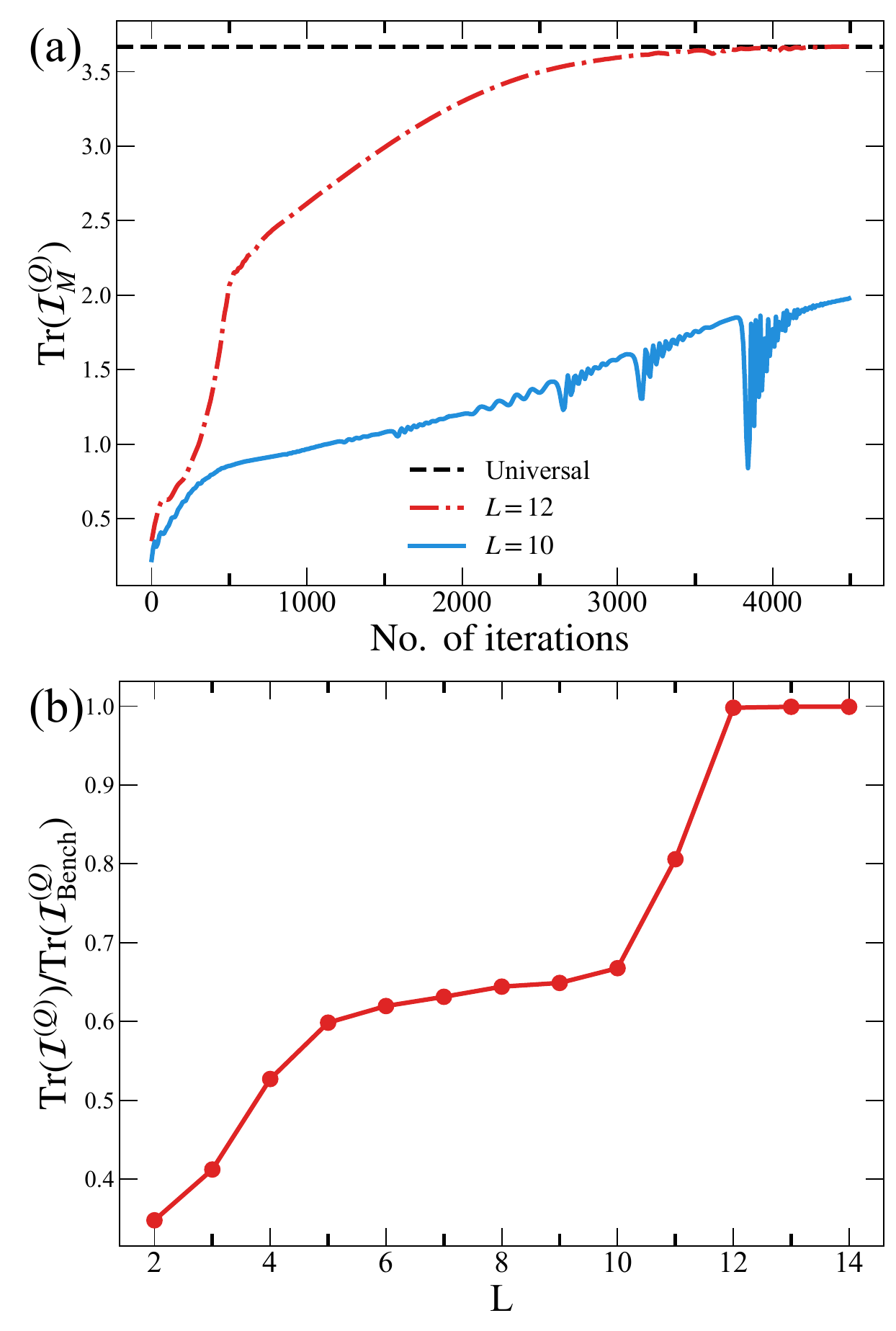}
	\caption{(a) Comparison of the sensing performance between the circuit discovered by \textsc{AutoQSense} and the universal ansatz, evaluated in terms of the cost function $\mathrm{Tr}(\mathcal{I}_M^{(Q)})$ for the 3D magnetic field estimation task. \textsc{AutoQSense} successfully matches the optimal performance, demonstrating its ability to autonomously discover high-precision sensing protocols. (b) The ratio of $\mathrm{Tr}(\mathcal{I}_M^{(Q)})$ to the optimal $\mathrm{Tr}(\mathcal{I}^{(Q)}_{\mathrm{Bench}})$ given by the universal two-qubit circuit.}
	\label{fig:two_qubit_comparison}
\end{figure}

Fig.~\ref{fig:two_qubit_comparison}(a) shows the value of $\mathrm{Tr}(\mathcal{I}^{(Q)}_M)$ as a function of number of iteration for representative gate budgets. For sufficiently large $L=12$, the learned architecture consistently converges to the benchmark QFI obtained from the universal circuit. In contrast, when the gate budget is restricted to $L=10$, $\mathrm{Tr}(\mathcal{I}^{(Q)}_M)$ reaches only half the value obtained with $L=12$ even after approximately 4,000 iterations, indicating that architectural capacity limits the accessible region of Hilbert space.

\paragraph{Convergence and expressive completeness.}

Fig.~\ref{fig:two_qubit_comparison}(b) shows, as a function of the gate budget, the ratio of the final saturated $\mathrm{Tr}(\mathcal{I}^{(Q)}_M)$ obtained by \textsc{AutoQSense} to that of the universal benchmark. We observe a sharp transition: at or below a critical architectural capacity around $L=11$, the achievable QFI is bounded away from the optimum, whereas beyond this threshold the learned architecture saturates the benchmark within numerical precision. 

Importantly, the training curves increase monotonically on average over independent random seeds, showing that the RL-based search remains stable in the discrete architecture space. With increasing gate budget, the saturated QFI systematically approaches the universal-circuit benchmark, and no degradation is observed once the search space becomes sufficiently expressive. Therefore, the residual performance gap at small $L$ should be attributed to the explicit gate-budget constraint rather than to an intrinsic limitation of the \textsc{AutoQSense} framework. This indicates that no additional performance loss was observed beyond the expressivity limit set by the allowed gate budgets.

\paragraph{Discussion.}

These results establish that \textsc{AutoQSense} is capable of recovering known optimal two-qubit sensing strategies when provided with adequate architectural resources. In this controlled setting, the algorithm automatically discovers preparation circuits that reproduce the maximal $\mathrm{Tr}(\mathcal{I}_M^{(Q)})$ achievable within the full two-qubit unitary manifold. At the same time, the gate-budget study highlights the fundamental role of architectural expressivity in variational quantum sensing: insufficient structural capacity, rather than parameter optimization failure, limits attainable precision.

\subsubsection{Noise-adaptive probes beyond GHZ and Ramsey protocols}\label{sec:ghz}
The preceding analysis showed that, in the two-qubit setting, \textsc{AutoQSense} can recover the sensitivity of a universal parametrized circuit once the gate budget is sufficiently expressive. We now move beyond this controlled expressivity benchmark and examine whether the learned architectures can compete with standard metrological protocols in a multi-qubit setting. Specifically, we benchmark \textsc{AutoQSense} against GHZ interferometry and Ramsey spectroscopy, which provide canonical reference strategies for entanglement-enhanced and unentangled sensing, respectively \cite{proctor2018multiparameter}. This comparison allows us to assess not only the achievable sensitivity of the learned circuits, but also how their structure adapts as dephasing noise changes the relative advantage of entangled probe states.

\paragraph{Sensing task.}
We consider a three-qubit magnetic-field sensing problem in which each qubit evolves under a local Hamiltonian
\begin{equation}
    H_i = \phi_z^{(i)} \sigma_z^{(i)},
\end{equation}
with $\phi_z^{(i)}$ denoting the strength of magnetic-field applied to qubit $i$. The estimation target is the average field
\begin{equation}
    \overline{\phi}_z = \frac{1}{3}(\phi_z^{(1)} + \phi_z^{(2)} + \phi_z^{(3)}),
\end{equation}
corresponding to collective phase accumulation under independent local interactions \cite{meyer2021variational}.

\paragraph{Reference protocols.}

\emph{GHZ interferometry} prepares the maximally entangled state
\begin{equation}
(|000\rangle + |111\rangle)/\sqrt{2},
\end{equation}
which accumulates a collective phase proportional to $\sum_i \phi_z^{(i)}$. In ideal, noiseless settings, GHZ states achieve enhanced phase sensitivity through coherent phase amplification. However, their reliance on global entanglement renders them highly susceptible to decoherence, as any local dephasing disrupts the accumulated collective phase. \emph{Ramsey spectroscopy}, by contrast, prepares each qubit in an independent superposition state and measures accumulated local phases. This protocol avoids multipartite entanglement and is generally more robust to noise, but does not exploit collective enhancement.
\begin{figure}
	\includegraphics[width=\linewidth]{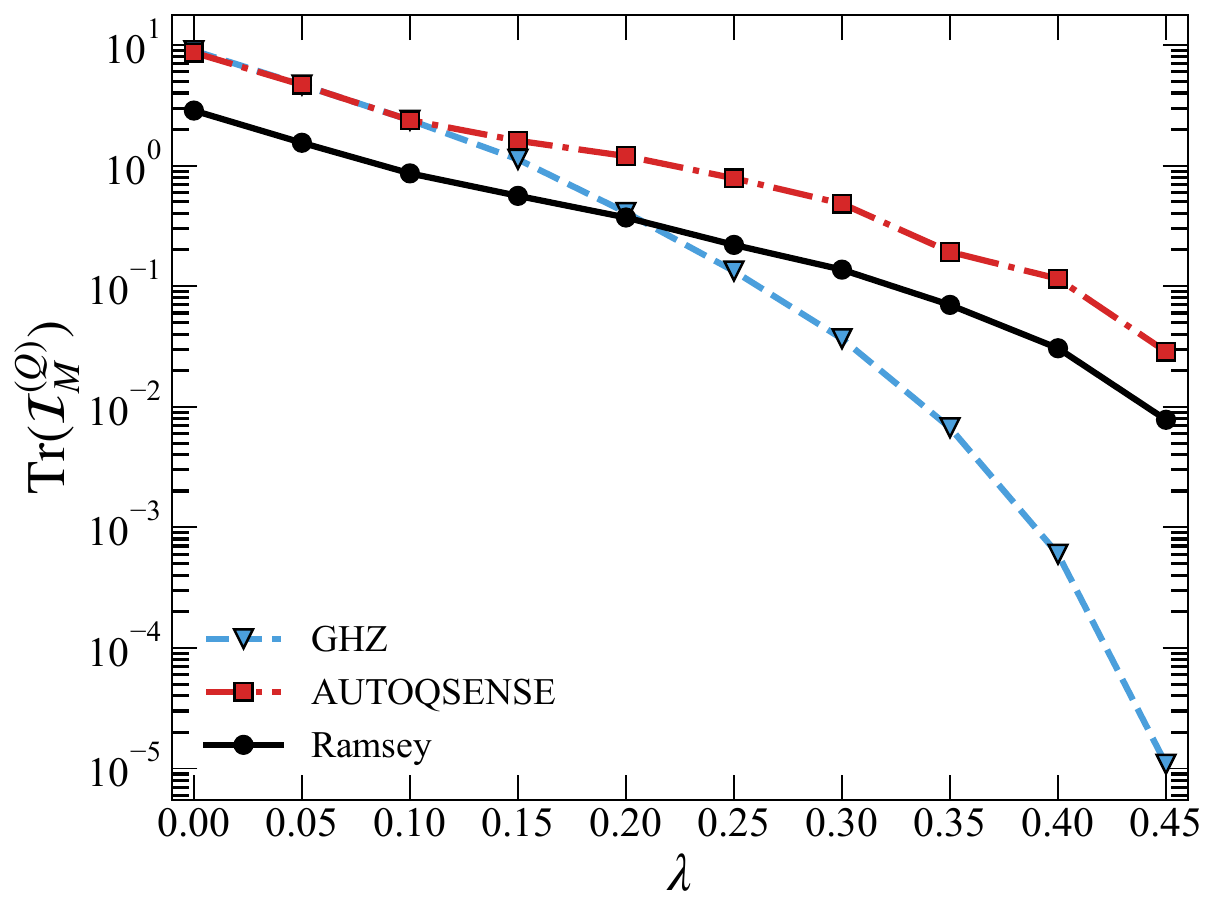}
	\caption{
Comparison of \textsc{AutoQSense} with GHZ interferometry and Ramsey spectroscopy for three-qubit magnetic-field sensing under dephasing noise of strength $\lambda$. $\mathrm{Tr}(\mathcal{I}^{(Q)}_M)$ (log scale) is plotted as a function of $\lambda$. In the low-noise regime, \textsc{AutoQSense} achieves performance comparable to GHZ interferometry and exceeds Ramsey spectroscopy. As dephasing increases, GHZ rapidly degrades due to its reliance on multipartite entanglement, while Ramsey exhibits gradual decay. \textsc{AutoQSense} maintains the highest $\mathrm{Tr}(\mathcal{I}^{(Q)}_M)$ across the entire noise range, demonstrating adaptive robustness to decoherence.}
\label{fig:ghz}
\end{figure}

\paragraph{Performance under dephasing noise.}

We introduce dephasing noise of strength $\lambda$ (see Eq.~\eqref{eq:dephasing}), acting on each qubit during the encoding process, and evaluate $\mathrm{Tr}(\mathcal{I}_M^{(Q)})$ as the performance metric. The results are shown in Fig.~\ref{fig:ghz}.

In the absence of noise, $\lambda=0$, both the GHZ protocol and \textsc{AutoQSense} achieve substantially higher sensitivity than Ramsey spectroscopy, reflecting the metrological advantage provided by multipartite entanglement. In this regime, the learned \textsc{AutoQSense} circuit prepares a probe state that is close to the GHZ state, showing that the architecture search can recover the known optimal entangled strategy when decoherence is absent. As the dephasing strength increases, the sensitivity of all three methods decreases. In the low-noise regime, $\lambda \lesssim 0.2$, GHZ and \textsc{AutoQSense} continue to outperform Ramsey spectroscopy, whereas for stronger dephasing, $\lambda \gtrsim 0.2$, the GHZ protocol degrades more rapidly and eventually falls below the Ramsey baseline due to the fragility of global coherence.

\begin{figure}
	\includegraphics[width=\linewidth]{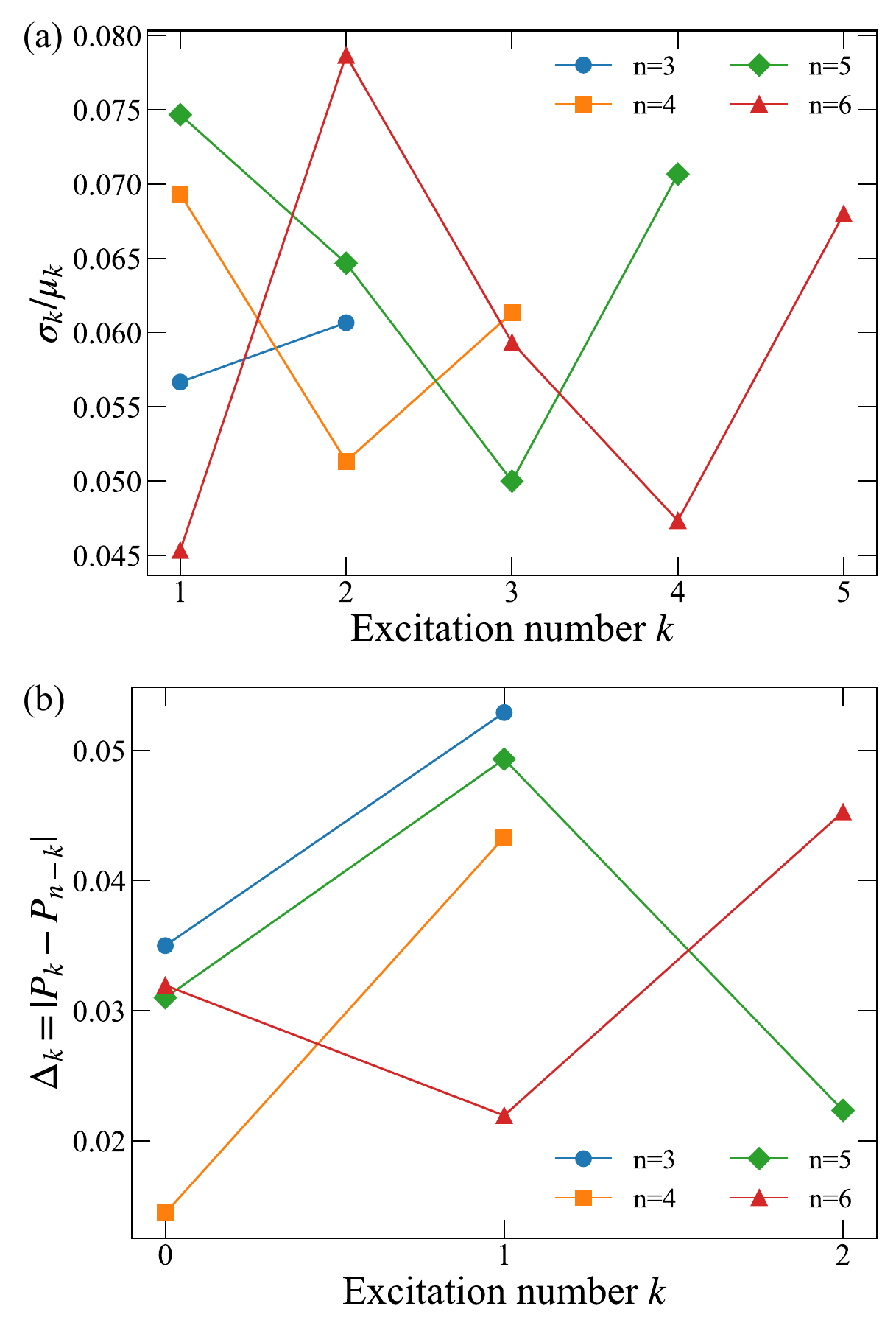}
	\caption{
Excitation-number symmetry and sector asymmetry of learned probe states. 
(a) Within-sector coefficient of variation, $\sigma_k/\mu_k$, for systems with $n=3,4,5,6$ qubits, computed across computational basis states with the same excitation number $k$, illustrating the uniformity of amplitudes within each sector. 
(b) Sector asymmetry, $\Delta_k = |P_k - P_{n-k}|$, quantifying the probability difference between complementary excitation-number sectors $k$ and $n-k$. 
}
	\label{fig:excitation_structure}
\end{figure}

Across the full range of noise strengths, however, \textsc{AutoQSense} consistently achieves higher Fisher information than both reference protocols. This advantage arises from its ability to adapt the probe-state structure to the noise level rather than committing to a fixed GHZ or Ramsey-type strategy. In the very low-noise regime, the learned state closely approximates the GHZ state, while under stronger dephasing it evolves toward a more robust, partially entangled state. The learned state also exhibits a clear excitation-number symmetry: computational basis states with the same number of excitations, such as $\ket{001}$, $\ket{010}$, and $\ket{100}$, acquire comparable probability amplitudes. This symmetry persists when the system size is extended to six qubits, suggesting a recurring structure rather than isolated circuit solutions. 

To quantify the structure of the learned probe state, we define the following quantities. Let $n$ be the total number of qubits, and for each computational basis state $\ket{z}$, let $w(z)$ denote its Hamming weight (i.e., the number of excitations or qubits in the $\ket{1}$ state). Denote the probability of observing basis state $\ket{z}$ as $p_z = |c_z|^2$,
where $c_z$ is the amplitude of $\ket{z}$ in the learned state $\ket{\psi} = \sum_{z\in\{0,1\}^n} c_z \ket{z}$. For a given excitation number $k \in \{0,1,\dots,n\}$, let $N_k = \binom{n}{k}$ be the number of basis states with Hamming weight $k$.

The mean probability within the $k$-excitation sector is
\begin{equation}
    \mu_k = \frac{1}{N_k}\sum_{z:w(z)=k} p_z,
\end{equation}
and the standard deviation within that sector is
\begin{equation}
    \sigma_k = \sqrt{\frac{1}{N_k}\sum_{z:w(z)=k}(p_z-\mu_k)^2}.
\end{equation}
The ratio $\sigma_k/\mu_k$ then quantifies the relative spread of probabilities among basis states with the same excitation number. Finally, the total weight of sector $k$ is
\begin{equation}
    P_k = \sum_{z:w(z)=k} p_z,
\end{equation}
and the asymmetry between complementary sectors is
\begin{equation}
    \Delta_k = |P_k - P_{n-k}|.
\end{equation}

Fig.~\ref{fig:excitation_structure} illustrates the structural properties of the probe states learned by \textsc{AutoQSense}. Panel (a) shows that the coefficient of variation $\sigma_k/\mu_k$ is consistently small across the dominant excitation sectors, indicating that basis states with the same number of excitations acquire comparable probabilities. Panel (b) shows the asymmetry $\Delta_k$ between complementary sectors, which remains small across all relevant sectors, confirming the approximate $k \leftrightarrow (n-k)$ symmetry. Together, these results suggest a recurring noise-adapted structure: amplitudes are relatively uniform within each excitation sector, and the total probability weight is nearly symmetric between sectors with complementary excitation numbers. This structure persists when the system size is increased from three to six qubits and under varying noise strengths, suggesting similar patterns across the simulated system sizes and noise settings.

\paragraph{Discussion.}

These results highlight two key observations. First, in the ideal regime, \textsc{AutoQSense} is capable of reproducing entanglement-enhanced sensing performance comparable to GHZ interferometry. Second, and more importantly, as noise increases, the learned circuits adapt their structure to mitigate decoherence-induced sensitivity loss. Unlike fixed analytic protocols, which are optimized for specific assumptions about noise, \textsc{AutoQSense} dynamically balances entanglement generation and robustness.

\subsubsection{Advantages over Fixed Hardware-Efficient Ans\"atze} \label{sec:defined_task}
\begin{figure}
	\includegraphics[width=\linewidth]{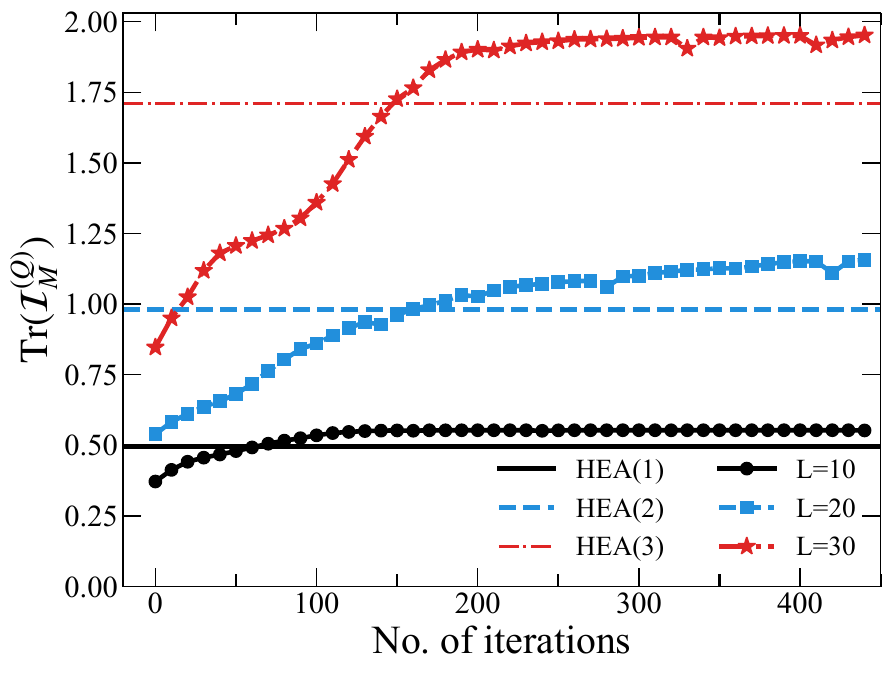}
	\caption{
Comparison between \textsc{AutoQSense} and hardware-efficient ans\"atze (HEA) in the multi-parameter sensing task defined in Sec.~\ref{sec:twoq_uni}. The curves show the training evolution of $\mathrm{Tr}(\mathcal{I}_M^{(Q)})$ as a function of optimization iterations for \textsc{AutoQSense} under different gate budgets $L=10,20,30$. The horizontal lines indicate the best performance achieved by HEA circuits with one, two, and three layers (HEA-1, HEA-2, HEA-3), trained to convergence.}
\label{fig:HEA_comparison_new}
\end{figure}

Having established that \textsc{AutoQSense} recovers the optimal solution in regimes where the exact optimum is known, we now investigate whether architectural adaptivity provides a tangible advantage in moderately sized systems where analytic solutions are unavailable and fixed-template VQS approaches are typically employed.
\begin{figure}
	\includegraphics[width=\linewidth]{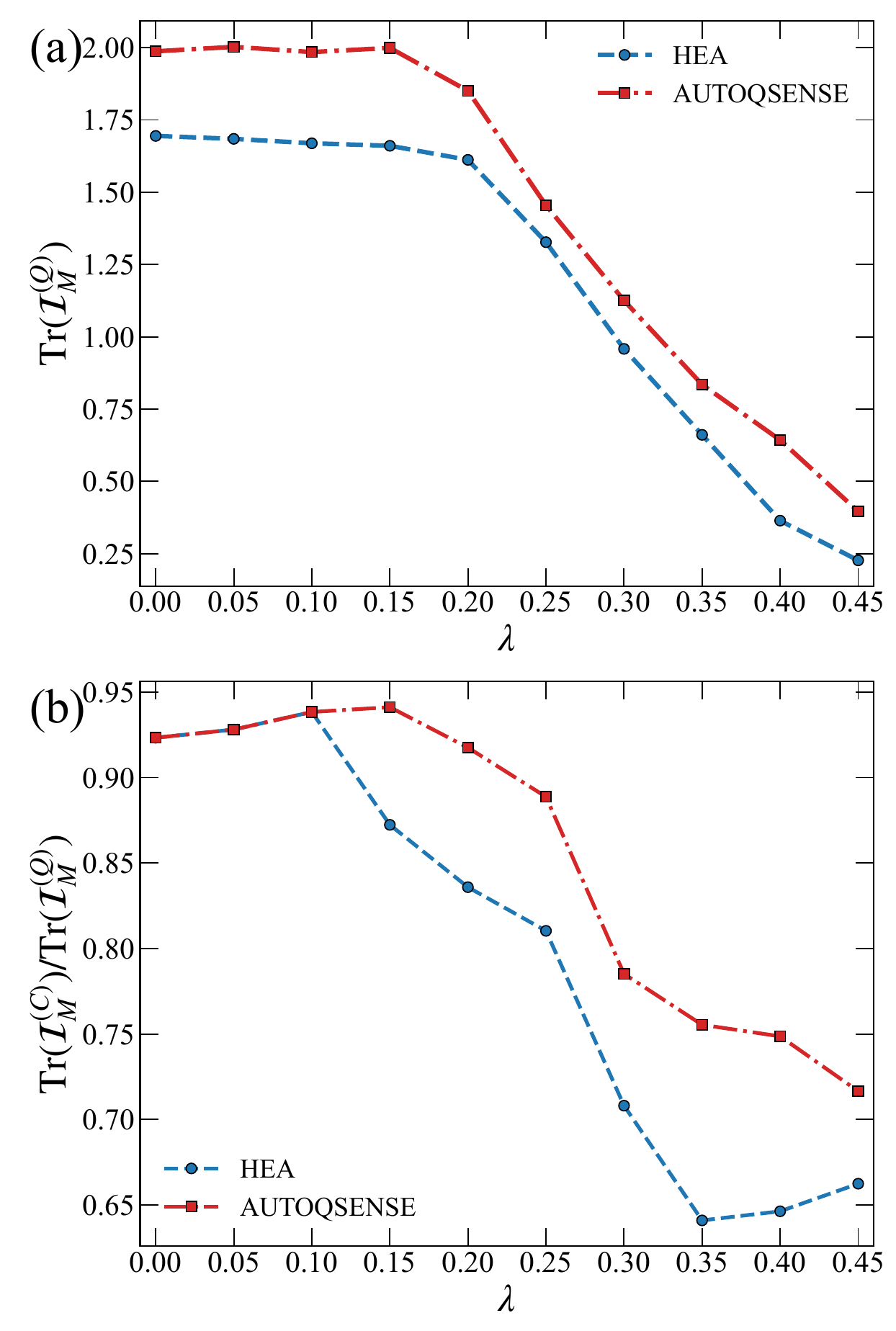}
	\caption{
Performance under dephasing noise for \textsc{AutoQSense} and HEA. 
(a) Trace of the quantum Fisher information matrix, $\mathrm{Tr}(\mathcal{I}_M^{(Q)})$, as a function of the dephasing strength $\lambda$. Both methods exhibit degradation with increasing noise, while \textsc{AutoQSense} consistently maintains higher sensitivity across the full range. 
	(b) Information-extraction ratio $\mathrm{Tr}(\mathcal{I}_M^{(C)})/\mathrm{Tr}(\mathcal{I}_M^{(Q)})$, a heuristic measure of how efficiently the implemented measurement extracts the available quantum Fisher information \textsc{AutoQSense} preserves a high information-extraction ratio even at strong noise, whereas HEA exhibits a pronounced decline, indicating reduced compatibility between probe and measurement under decoherence.}
	\label{fig:attainability}
\end{figure}

\paragraph{Benchmark against hardware-efficient ansatz.}

We benchmark \textsc{AutoQSense} against a family of HEA circuits, each consisting of alternating layers of parametrized single-qubit rotations and CNOT entangling gates arranged in a fixed connectivity pattern, under the same sensing task in Sec.~\ref{sec:twoq_uni}. We consider HEA circuits with one, two, and three layers, denoted HEA-1, HEA-2, and HEA-3, corresponding to progressively increasing parameter counts and entangling depth. 

\begin{figure*}[t]
	\includegraphics[width=\textwidth]{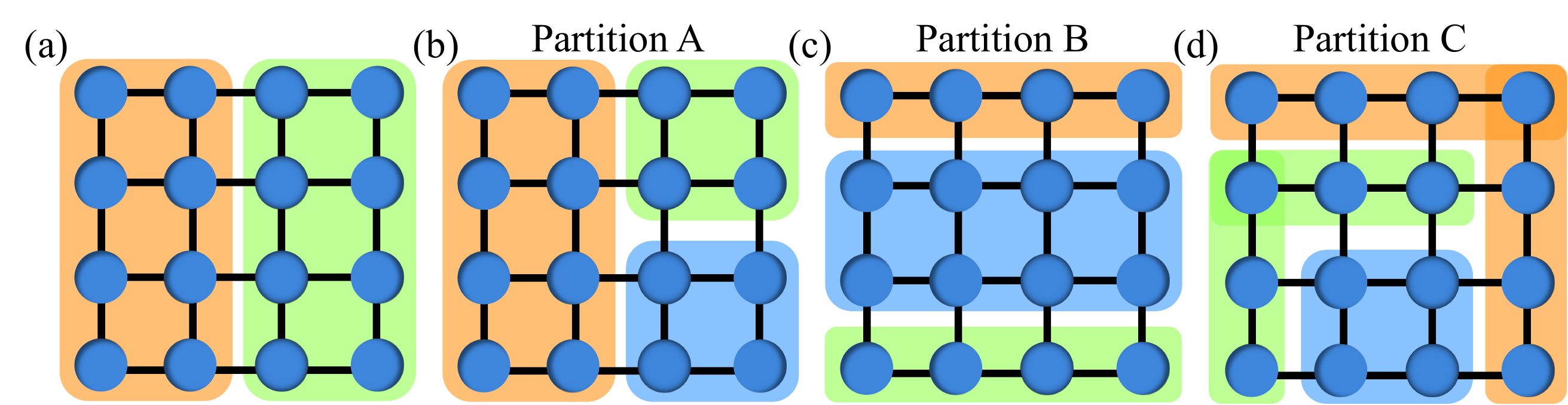}
	\caption{(a) Illustration of one possible partition of a $4 \times 4$ qubit array into two subsystems. Panels (b)--(d) illustrate three different partitioning schemes (A, B, and C) for the distributed reinforcement learning setup with three local agents and one entangling agent. Local agents control intra-block actions, and an entangling agent manages inter-block interactions. The connections between blocks represent possible entangling operations in the system.}
	\label{fig:partition_illustrate}
\end{figure*}

To assess expressivity in the multi-parameter sensing regime, we evaluate $\mathrm{Tr}(\mathcal{I}_M^{(Q)})$ to examine the probe state produced by the preparation circuit. The parameters in HEA circuits are trained under optimization conditions identical to those used for \textsc{AutoQSense}, ensuring a controlled comparison at matched circuit depths.

Fig.~\ref{fig:HEA_comparison_new} shows the training trajectories of \textsc{AutoQSense} for several gate budgets $L$, with dashed horizontal lines indicating the best performance achieved by HEA at the corresponding depth. Across all gate budgets, \textsc{AutoQSense} attains higher $\mathrm{Tr}(\mathcal{I}_M^{(Q)})$, and the performance gap increases with circuit depth. This behavior indicates that additional circuit capacity is exploited more effectively when the architecture itself is optimized. In contrast, the fixed HEA layout fails to translate increased depth into commensurate gains in Fisher information, revealing structural limitations inherent to manually designed templates.

\paragraph{Robustness under noisy encoding.}

We next examine performance under dephasing noise of strength $\lambda \in [0,0.45]$ acting during the encoding dynamics. $\mathrm{Tr}(\mathcal{I}_M^{(Q)})$ remains our primary performance metric. As shown in Fig.~\ref{fig:attainability}(a), both approaches exhibit the expected monotonic degradation of $\mathrm{Tr}(\mathcal{I}_M^{(Q)})$ with increasing noise strength, reflecting the loss of coherence in the probe state. However, \textsc{AutoQSense} consistently maintains a higher $\mathrm{Tr}(\mathcal{I}_M^{(Q)})$ throughout the entire noise range.

\paragraph{Measurement information extraction.}

To evaluate measurement performance, we compute the ratio
$
\mathrm{Tr}(\mathcal{I}^{(C)}_M)/\mathrm{Tr}(\mathcal{I}_M^{(Q)})$,
which serves as a heuristic information-extraction ratio for the jointly optimized preparation and measurement circuits. A ratio approaching unity indicates more efficient extraction of the available quantum Fisher information.

As shown in Fig.~\ref{fig:attainability}(b), HEA exhibits a pronounced decline in information extraction as dephasing increases, indicating that its fixed readout architecture becomes suboptimal in noisy regimes. In contrast, \textsc{AutoQSense} maintains ratios above $0.7$ across the entire noise range, demonstrating that the co-optimization of preparation and measurement preserves sensing performance even under substantial noise.

\paragraph{Summary.}

These results highlight that (i) structural adaptivity provides substantial expressive benefits, (ii) robustness to noise emerges naturally from the learned circuit architecture, and (iii) effective measurement strategies are better captured by adaptive design.

\subsection{AutoQSense in Large-Scale Quantum Sensing Using a Multi-Agent Framework}

\begin{table*}
\centering
\begin{tabular}{cccc}
\hline
\textbf{Partition} & \textbf{Single-Qubit Rotations} & \textbf{Two-Qubit Interactions} & \textbf{Entangling Agent Actions} \\ \hline
\textbf{Partition A} & (8,4,4) & (10,4,4) & (2,2,2) \\ 
\textbf{Partition B} & (4,4,8) & (3,3,10) & (0,4,4) \\ 
\textbf{Partition C} & (7,5,4) & (6,4,4) & (3,2,4) \\ 
\textbf{No Partition} & 16 & 24 & 0 \\
\hline
\end{tabular}
\caption{Summary of the action-space sizes for each partition scheme in the distributed reinforcement learning setup, together with the corresponding action count for the unpartitioned system. The first three rows list the numbers of single-qubit rotations, intra-block two-qubit interactions, and entangling-agent actions for each partition configuration (A, B, and C). Entries for the first two action types are ordered by subsystem color (orange, green, blue), while entangling-agent actions are ordered by inter-block connection type: (orange, green), (orange, blue), and (green, blue). The last row lists the total numbers of single-qubit rotations and two-qubit interactions for the unpartitioned system.}
\label{tab:count}
\end{table*}
The few-qubit results demonstrate that \textsc{AutoQSense} can recover optimal or near-optimal sensing structures, adapt to noise, and outperform fixed-structure ans\"atze when the architecture search space is sufficiently expressive. However, directly extending the single-agent formulation to larger systems becomes increasingly inefficient, because the number of possible circuit structures grows rapidly with both the number of qubits and the gate budget. Moreover, in realistic large-scale devices, circuit design is often constrained by modular connectivity and limited entangling operations between distant qubits. These considerations motivate the multi-agent version of \textsc{AutoQSense}, which decomposes the global architecture search into coordinated local decisions while retaining the ability to introduce inter-block entanglement through a restricted interface. We next evaluate distributed-\textsc{AutoQSense} on larger sensing tasks and examine how subsystem partitioning affects the achievable CFI. We use the sensing task defined in Sec.~\ref{sec:ghz} with $\lambda=5\times10^{-3}$. The CFI serves as the optimization objective, while the QFI is reported separately to assess probe-state quality.

To enable scalable circuit design for large quantum systems, we employ a subsystem-based partition of the full device and formulate the circuit construction problem within a multi-agent reinforcement learning framework. Fig.~\ref{fig:partition_illustrate} illustrates the partitioning scheme for a 16-qubit system arranged in a $4\times4$ lattice with nearest-neighbor connectivity. In Fig.~\ref{fig:partition_illustrate} (a), the system is divided into two subsystems, while Fig.~\ref{fig:partition_illustrate} (b)--(d) present three distinct three-block partition configurations. Colored regions denote subsystems controlled by local agents that are restricted to applying intra-block single- and two-qubit gates, with available operations determined by the hardware.
\begin{figure} 
	\includegraphics[width=\linewidth]{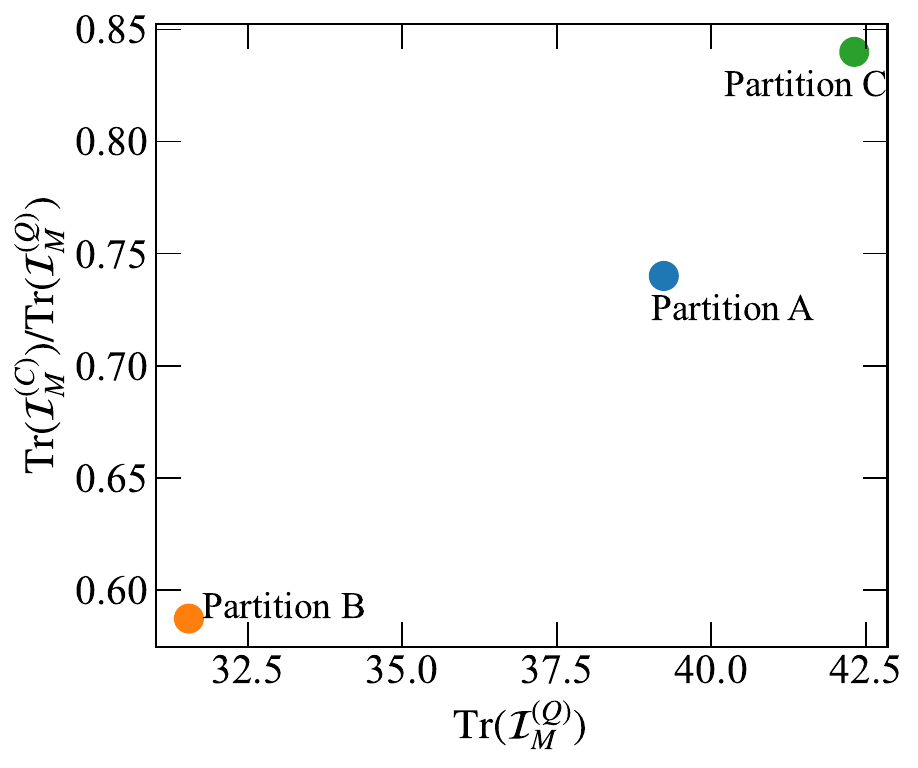}
	\caption{
Comparison of different subsystem partitioning strategies in the multi-agent architecture search framework. Each point corresponds to a partition configuration and is positioned according to the achieved $\mathrm{Tr}(\mathcal{I}_M^{(Q)})$ and the measurement efficiency quantified by the ratio $\mathrm{Tr}(\mathcal{I}_M^{(C)})/\mathrm{Tr}(\mathcal{I}_M^{(Q)})$. Partitions located toward the upper-right corner simultaneously exhibit stronger metrological sensitivity and more efficient extraction of the encoded information through the learned measurement protocol.
}
	\label{fig:partition_comparison}
\end{figure}

\begin{figure}
	\includegraphics[width=\linewidth]{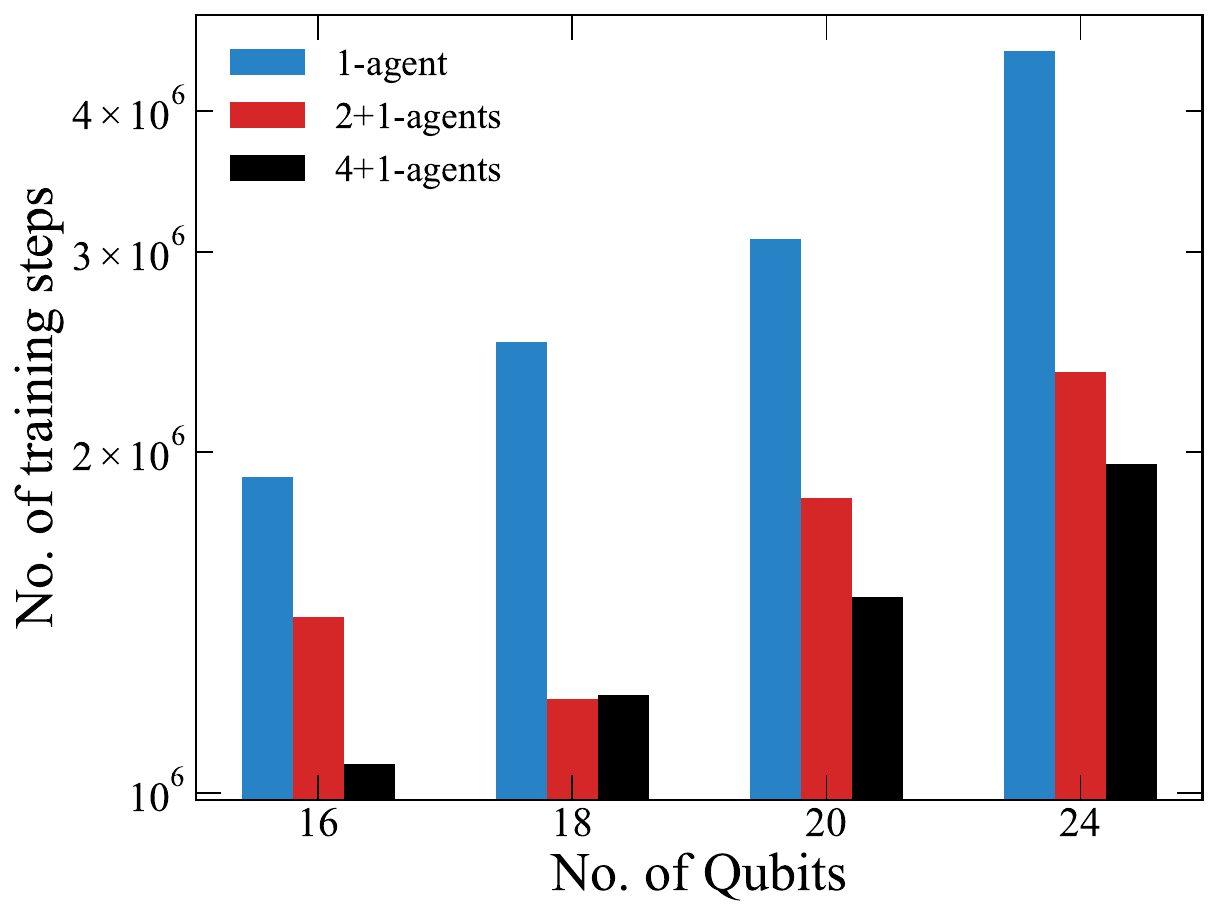}
	\caption{Number of training steps required for \textsc{AutoQSense} to reach the HEA level of $\mathrm{Tr}(\mathcal{I}^{(C)}_{M})$ for different multi-agent configurations.}
	\label{fig:large_comparison}
\end{figure}
Inter-block entanglement is mediated by a dedicated entangling agent, which is constrained to apply a limited number of two-qubit gates across subsystem boundaries. This architectural design enforces locality while still enabling the generation of global correlations required for metrological performance. The corresponding action-space sizes for each agent under different partitioning schemes are summarized in Table~\ref{tab:count}. Notably, partitioning reduces the local action space per agent compared to the unpartitioned setting (from 40 total actions to distributed subsets), thereby decreasing the effective search complexity.
\begin{table*} 
\centering 
\begin{tabular}{ccc| |cc| |cc| |cc} 
\toprule 
\hline 
\multirow{2}{*}{\# Qubits $n$} & \multirow{2}{*}{\# Blocks $B$} & \multirow{2}{*}{Gate budget $L$} & \multicolumn{2}{c}{$\mathrm{Tr}(\mathcal{I}^{(Q)}_M)$} & \multicolumn{2}{c}{$\mathrm{Tr}(\mathcal{I}^{(C)}_M)/\mathrm{Tr}(\mathcal{I}^{(Q)}_M)$} & \multicolumn{2}{c}{\# 2Q gates $N_{2q}$} \\ 
\cmidrule(lr){4-5} \cmidrule(lr){6-7} \cmidrule(lr){8-9} & & & \textsc{AutoQSense} & HEA & \textsc{AutoQSense} & HEA & \textsc{AutoQSense} & HEA \\ 
\hline 
16 & 2 & 240 & 36.63 & 21.04 & 0.81 & 0.85 & 84 & 128 \\ 
16 & 4 & 240 & 40.11 & 21.04 & 0.76 & 0.85 & 91 & 128 \\ 
18 & 2 & 300 & 46.04 & 23.62 & 0.72 & 0.74 & 89 & 144 \\ 
18 & 4 & 300 & 48.71 & 23.62 & 0.77 & 0.74 & 71 & 144 \\ 
20 & 2 & 300 & 54.75 & 30.85 & 0.83 & 0.71 & 98 & 160 \\ 
20 & 4 & 300 & 59.67 & 30.85 & 0.88 & 0.71 & 88 & 160 \\ 
24 & 2 & 300 & 71.88 & 43.28 & 0.76 & 0.66 & 121 & 192 \\ 
24 & 4 & 300 & 83.69 & 43.28 & 0.77 & 0.66 & 102 & 192 \\ \hline

\bottomrule

\end{tabular}
\caption{
Comparison of quantum and classical Fisher information metrics for various numbers of qubits ($n$), circuit block counts ($B$), and gate budgets ($L$) using the \textsc{AutoQSense} ansatz versus a hardware-efficient ansatz (HEA). The table reports the trace of the quantum Fisher information matrix $\mathrm{Tr}(\mathcal{I}^{(Q)}_M)$, the ratio of classical-to-quantum Fisher information $\mathrm{Tr}(\mathcal{I}^{(C)}_M)/\mathrm{Tr}(\mathcal{I}^{(Q)}_M)$, and the total number of two-qubit gates $N_{2q}$ for each configuration.
}
\label{tab:results_summarize}
\end{table*}

We emphasize that the choice of partitioning has a direct impact on both learning efficiency and circuit performance. In particular, partitioning strategies that preserve the greatest degree of physical connectivity between subsystems yield superior results. As shown in Fig.~\ref{fig:partition_comparison}, Partition A and Partition C outperform Partition B. This performance gap can be attributed to the absence of direct interactions between the orange and green subsystems in Partition B, which restricts the ability of the entangling agent to establish correlations. Consequently, the achievable performance is reduced in Partition B relative to configurations that maintain inter-block connectivity.

Using this decomposition, we perform numerical benchmarks of \textsc{AutoQSense} against the HEA baseline for system sizes ranging from 16 to 24 qubits. The reward is defined in Eq.~\eqref{eq:diff-reward}.

In Fig.~\ref{fig:large_comparison}, we report the number of training steps required for \textsc{AutoQSense} to reach the performance of the HEA baseline. Multi-agent configurations consistently reduce the required training steps compared to the single-agent setting. This reduction becomes more pronounced as the system size increases, indicating improved scaling behavior. 

\begin{figure}[t]
	\includegraphics[width=\linewidth]{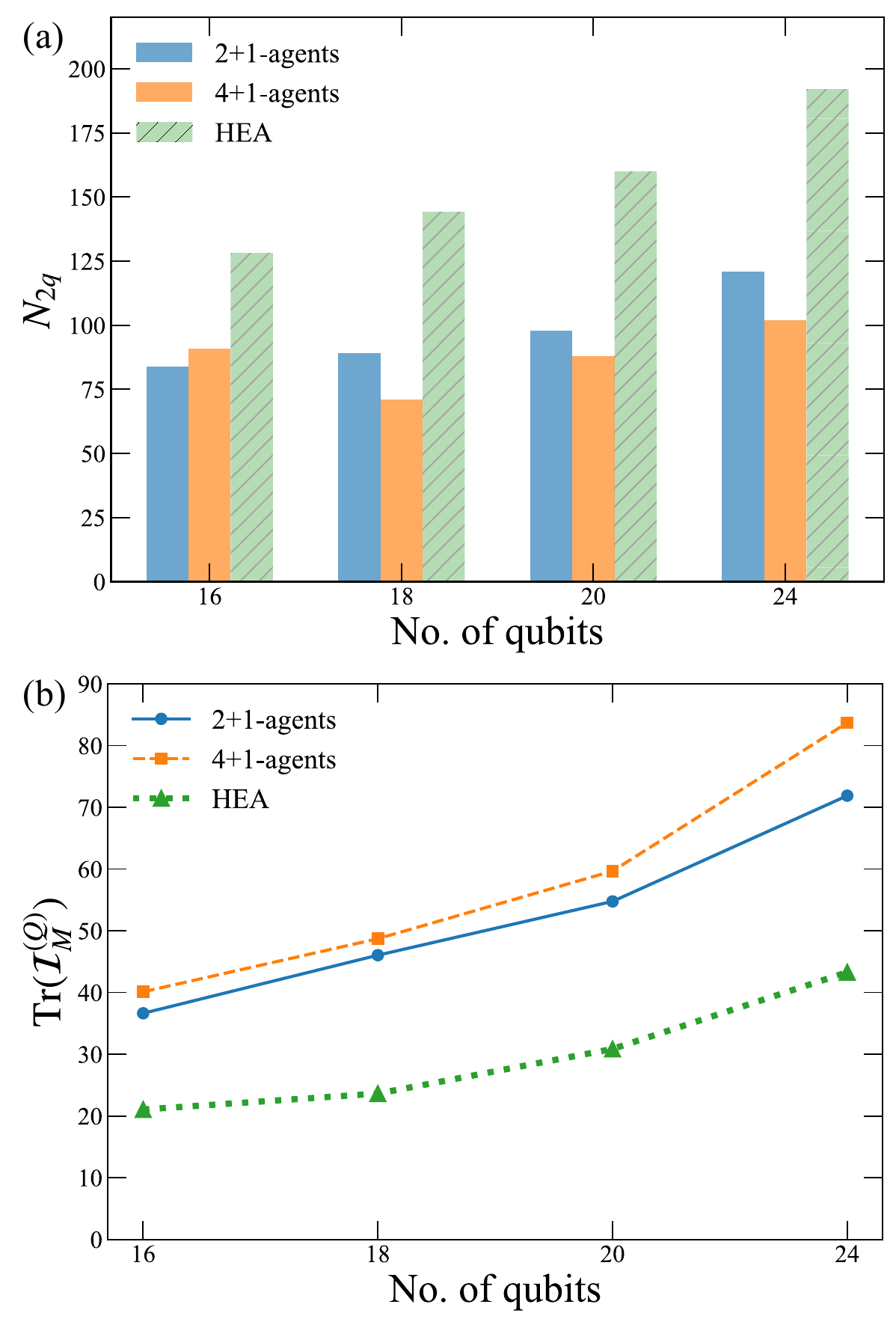}
	\caption{Comparison of circuit resources and metrological performance for multi-agent \textsc{AutoQSense} and HEA circuits. (a) Number of two-qubit gates, $N_{2q}$, for the three-agent and five-agent distributed architectures and the HEA baseline. (b) $\mathrm{Tr}(\mathcal{I}^{(Q)}_M)$ for the same architectures as a function of the number of qubits.}
	\label{fig:dual_axis}
\end{figure}

Table~\ref{tab:results_summarize} and Fig.~\ref{fig:dual_axis} further show that \textsc{AutoQSense} achieves substantial improvements in the QFI-matrix trace, $\mathrm{Tr}(\mathcal{I}^{(Q)}_M)$. For instance, for 24 qubits with 4 blocks, the value increases from 43.28 for HEA to 83.69 for \textsc{AutoQSense}, corresponding to an improvement of approximately 93.4\%. Similar improvements are observed across all system sizes, with gains consistently exceeding 50\%.

The information-extraction ratio $\mathrm{Tr}(\mathcal{I}^{(C)}_M)/\mathrm{Tr}(\mathcal{I}^{(Q)}_M)$ remains comparable in some settings and improves in others. \textsc{AutoQSense} achieves comparable or superior performance while significantly reducing the number of entangling gates.

Specifically, for 16 qubits, the two-qubit gate count is reduced from 128 (HEA) to as low as 84, corresponding to a reduction of approximately 34\%. Similarly, for 24 qubits, the gate count is reduced from 192 to 102, yielding a reduction of approximately 46\%. This consistent reduction in $N_{2q}$ demonstrates that the learned circuits utilize entangling operations more efficiently by placing them only where they contribute most to performance, rather than following a fixed layered structure.

\begin{figure}
	\includegraphics[width=\linewidth]{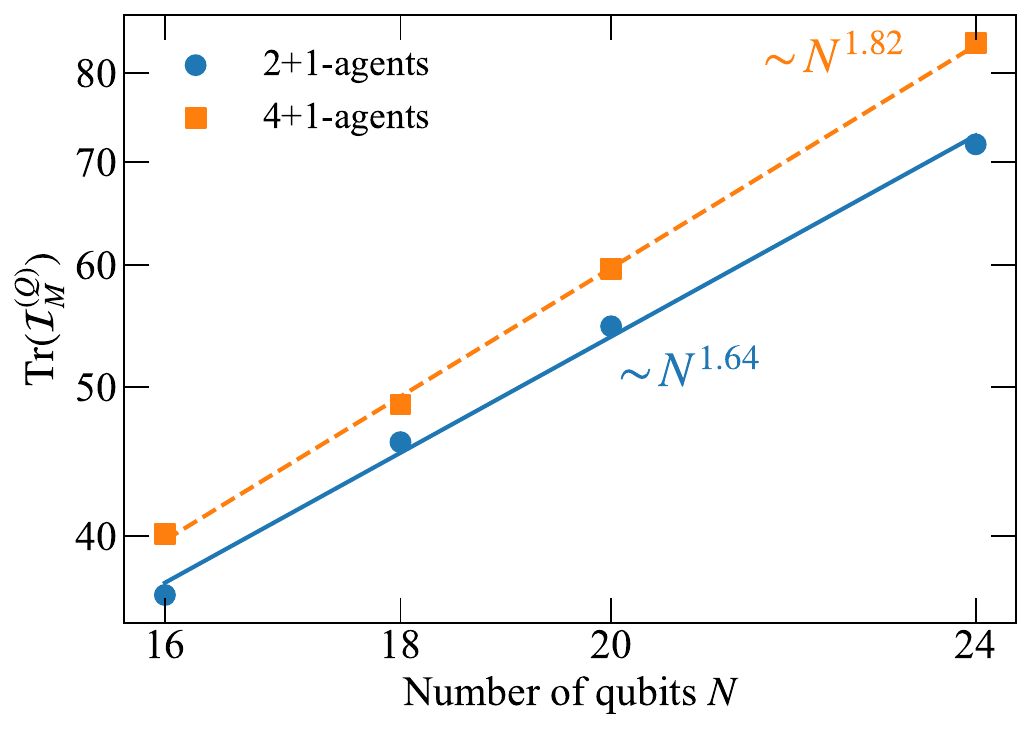}
	\caption{Scaling of the quantum Fisher information with system size for distributed-\textsc{AutoQSense}. Markers show the optimized values of $\mathrm{Tr}(\mathcal{I}_M^{(Q)})$ obtained for subsystem partitions with $2+1$ agents and $4+1$ agents, while solid curves represent power-law fits of the form $\mathrm{Tr}(\mathcal{I}_M^{(Q)})\sim N^{\ell}$.
}
	\label{fig:scaling_qfi}
\end{figure}

Fig.~\ref{fig:scaling_qfi} further examines how the optimized sensing performance scales with system size for different subsystem partitioning strategies. The markers denote the values of $\mathrm{Tr}(\mathcal{I}^{(Q)}_M)$ obtained from the learned architectures, while the solid lines show power-law fits of the form $\mathrm{Tr}(\mathcal{I}^{(Q)}_M)=aN^{\ell}$. For both partition choices, $\mathrm{Tr}(\mathcal{I}^{(Q)}_M)$ increases monotonically with the number of qubits, indicating that the distributed architecture-search framework remains effective as the sensing system grows. Moreover, the $4+1$-agent partition consistently achieves larger values of $\mathrm{Tr}(\mathcal{I}^{(Q)}_M)$ than the $2+1$-agent partition across all investigated system sizes, with the performance gap becoming more pronounced for larger systems.

The fitted scaling exponents provide additional insight into the role of subsystem decomposition. A larger number of blocks increases the granularity of the architecture search and allows the learning agents to explore a richer set of local sensing structures while coordinating through a smaller set of inter-block entangling operations. As a result, the $4+1$-agents configuration exhibits a more favorable scaling trend than the $2+1$-agents configuration, suggesting that an appropriate partition strategy can improve both search efficiency and achievable metrological performance. Although the present system sizes are not sufficient to establish asymptotic scaling laws, the observed behavior demonstrates that the distributed-\textsc{AutoQSense} framework can maintain its advantage across the tested system sizes.

These results demonstrate that subsystem partitioning combined with multi-agent reinforcement learning improves both training efficiency and circuit quality. The proposed approach reduces the number of required training steps, decreases circuit complexity through a significant reduction in two-qubit gates, and improves metrological performance. These advantages become increasingly pronounced with system size, demonstrating improved performance over the tested range relative to conventional HEA constructions.

\section{Conclusion and Outlook}
\label{sec:conclusion}

In this work, we introduced \textsc{AutoQSense}, an automated framework for designing variational quantum sensing protocols through reinforcement-learned circuit structures. Rather than assuming a fixed parametrized ansatz, \textsc{AutoQSense} treats the preparation and measurement circuits as architecture-level optimization variables and searches over physically realizable gate sequences using metrological objective functions. This formulation reframes variational quantum sensing as a sequential decision-making problem, in which the reward is directly tied to the quantum or classical Fisher information associated with the sensing task.

Our numerical results show that architecture search provides clear advantages over fixed-structure design. In the two-qubit benchmark, \textsc{AutoQSense} recovers the sensitivity achieved by a universal parametrized circuit once the gate budget becomes sufficiently expressive, indicating that the RL formulation does not introduce additional variational bias beyond explicit depth constraints. In multi-qubit sensing under dephasing noise, the learned circuits reproduce GHZ-like strategies in the ideal limit while adapting toward more robust partially entangled probe states as noise increases. Compared with both Ramsey and GHZ protocols, \textsc{AutoQSense} maintains higher Fisher information across the tested noise regimes. Moreover, against hardware-efficient ans\"atze under fixed resource settings, the learned architectures consistently achieve stronger sensitivity and capture a larger fraction of the available quantum Fisher information.

To address the rapid growth of the architecture search space, we further developed a distributed multi-agent version of \textsc{AutoQSense}. By decomposing the global circuit into locally controlled subsystems and introducing inter-block entanglement through a restricted interface, the multi-agent formulation reduces the effective search complexity while preserving the ability to generate useful multipartite correlations. Our large-scale results show that this modular strategy remains competitive beyond the few-qubit regime and that the choice of subsystem partition can affect the final classical Fisher information. These findings highlight the importance of architecture-aware and hardware-compatible design in scalable quantum sensing.

Beyond performance improvements, the learned circuits provide qualitative insight into the structure of noise-adapted sensing protocols. In particular, the optimized probe states exhibit organized amplitude patterns across excitation-number sectors, suggesting that \textsc{AutoQSense} does not merely discover isolated high-performing circuits, but can reveal recurring structural features associated with robust metrological performance. This interpretability is important because automated design tools should not only improve numerical benchmarks, but also help identify principles that can guide future sensing protocols.

Several directions remain open for future work. First, the present framework can be extended to more realistic noise models, including correlated dephasing, amplitude damping, crosstalk, and device-specific calibration errors. Second, the reward function can be generalized beyond the current Fisher-information-based objectives to include explicit resource costs such as circuit depth, two-qubit gate count, measurement overhead, or robustness to model uncertainty. Third, the multi-agent formulation can be further improved through more refined credit-assignment mechanisms, curriculum learning, or hierarchical policies that explicitly coordinate local and inter-block decisions. Finally, implementing \textsc{AutoQSense} on experimental quantum hardware would provide an important test of whether reinforcement-learned architectures can deliver practical sensing advantages under finite sampling, imperfect gates, and constrained connectivity.

Overall, we established reinforcement-learned architecture discovery as a systematic tool for quantum metrology. By elevating circuit structure to a trainable object and directly optimizing task-specific sensing objectives, \textsc{AutoQSense} provides a pathway toward adaptive, noise-aware, and hardware-compatible quantum sensing protocols for increasingly complex quantum devices.

\section*{ACKNOWLEDGEMENTS}
This work is supported by the National Natural Science Foundation of China (Grant No. 12474489), the Shenzhen Fundamental Research Program (Grant No. JCYJ20240813153139050), and the Guangdong Provincial Quantum Science Strategic Initiative (Grant No. GDZX2203001, GDZX2403001). The calculations involved in this work were partially performed on the Tianhe Xingyi supercomputer at the National Supercomputer Center in Guangzhou, China.

\end{document}